\documentclass[10pt,conference]{IEEEtran}
\usepackage{graphicx}
\usepackage[ruled,vlined,linesnumbered,noend]{algorithm2e}
\usepackage{figlatex,wrapfig}
\usepackage{listings,amssymb}
\usepackage{mathtools}
\usepackage{stmaryrd}
\usepackage{mathrsfs}
\usepackage{array,multirow}
\usepackage{caption}
\usepackage{wrapfig}
\usepackage{textcomp}
\usepackage[hidelinks]{hyperref}
\usepackage[dvipsnames]{xcolor}
\usepackage{xspace}
\usepackage{amsmath}
\usepackage{afterpage}
\usepackage{stackengine}
\usepackage{import}
\usepackage{float}
\usepackage[normalem]{ulem}
\hypersetup{
	colorlinks = true,
	citecolor = {magenta},
	linkcolor = {purple},
	urlcolor  = {blue}
}
\usepackage{tikz}
\usetikzlibrary{positioning}
\usetikzlibrary{shapes}
\usetikzlibrary{shapes,fit,backgrounds}
\usetikzlibrary{calc}

\newcommand\redsout{\bgroup\markoverwith{\textcolor{red}{\rule[0.5ex]{2pt}{0.4pt}}}\ULon}

\newcommand{\ext}[1]{\textcolor{PineGreen}{#1}}

\newcounter{defcounter}

\newcounter{lemmacounter}
\newcommand{\lemma}[2]{
	
	\medskip
	\refstepcounter{lemmacounter}
	
	\noindent
	\textit{\textbf{Lemma \arabic{lemmacounter}}. #1}
	
	\label{lem:#2}
	\medskip
}

\newcounter{theoremcounter}
\newcommand{\theorem}[2]{
	
	\refstepcounter{theoremcounter}
	
	\noindent\newline 
	\textit{\textbf{Theorem \arabic{theoremcounter}}. #1}
	
	\label{thm:#2}
}

\newcommand{\proof}[1]{\noindent\textit{Proof.} #1}

\newcommand{\hl}[1]{\hypertarget{#1}{\textcolor{darkgray}{\texttt{#1}}}\xspace} 
\newcommand{\hlref}[1]{\hyperlink{#1}{\textcolor{Sepia}{\small \texttt{#1}}}}
\newcommand{\tab}{\quad\quad}

\newcommand{\refsection}[1]{$\mathsection${\ref{#1}}\xspace}

\newcommand{\reffig}[1]{Fig.~\ref{#1}\xspace}
\newcommand{\reftab}[1]{Table~\ref{#1}\xspace}

\renewcommand{\iff}{\xspace\Leftrightarrow\xspace}
\renewcommand{\implies}{\xspace\Rightarrow\xspace}

\newcommand{\ie}{{\em i.e.}\xspace}
\newcommand{\st}{\ \mbox{s.t.}\ }
\newcommand{\viz}{\textit{viz}.\@\xspace}
\newcommand{\wrt}{\textit{wrt}\xspace}

\newcommand{\sota}{state-of-the-art\xspace}

\renewcommand{\^}{\xspace\wedge\xspace}
\renewcommand{\v}{\xspace\vee\xspace}

\renewcommand{\|}{\ |\ }

\newcommand{\union}{\xspace\cup\xspace}

\newcommand{\nin}{\not\in\xspace}

\newcommand{\cmodel}{\textsl{C11}\xspace}
\newcommand{\mca}{\textsl{MCA}\xspace}
\newcommand{\rinspect}{rInspect\xspace}
\newcommand{\ourtechnique}{\textcolor{Sepia}{\texttt{MoCA}}\xspace}
\newcommand{\ourtool}{\ourtechnique{-}tool\xspace}
\newcommand{\pretransformation}{early-write\xspace}
\newcommand{\Pretransformation}{Early-write\xspace}
\newcommand{\pretransformed}{early-write\xspace}
\newcommand{\sdpor}{source-DPOR\xspace}
\newcommand{\Sdpor}{Source-DPOR\xspace}
\newcommand{\cds}{CDSChecker\xspace}
\newcommand{\genmc}{GenMC\xspace}
\newcommand{\hmc}{HMC\xspace}

\newcommand{\program}{$P$\xspace}
\newcommand{\programhat}{$\widehat{P}$\xspace}
\newcommand{\states}{\Sigma\xspace}
\newcommand{\memords}{\mathcal{M}\xspace}
\newcommand{\actions}{\mathcal{A}\xspace}
\newcommand{\objects}{\mathcal{O}\xspace}
\newcommand{\s}[1]{s_{[#1]}\xspace} 
\newcommand{\p}[1]{P_{#1}\xspace}

\newcommand{\wt}[1]{{\mathbb{W}#1}}
\newcommand{\md}[1]{{\mathbb{M}#1}}
\newcommand{\rd}[1]{{\mathbb{R}#1}}
\newcommand{\fn}[1]{{\mathbb{F}#1}}
\newcommand{\events}{\mathcal{E}\xspace}
\newcommand{\writes}{\events^\wt{}\xspace}
\newcommand{\modifiers}{\events^\md{}\xspace}

\newcommand{\reads}{\events^\rd{}\xspace}
\newcommand{\fences}{\events^\fn{}\xspace}
\newcommand{\ordevents}[1]{\events^{(#1)}\xspace}
\newcommand{\ordwrites}[1]{\events^\wt{(#1)}}
\newcommand{\ordmodifiers}[1]{\events^\md{(#1)}}

\newcommand{\ordreads}[1]{\events^\rd{(#1)}}
\newcommand{\ordfences}[1]{\events^\fn{(#1)}}

\newcommand{\na}{\texttt{na}\xspace}
\newcommand{\rlx}{\texttt{rlx}\xspace}
\newcommand{\rel}{\texttt{rel}\xspace}
\newcommand{\acq}{\texttt{acq}\xspace}
\newcommand{\acqrel}{\texttt{acq-rel}\xspace}
\renewcommand{\sc}{\texttt{sc}\xspace}

\newcommand{\equi}{\sim\xspace}
\newcommand{\nequi}{\nsim\xspace}
\newcommand{\molt}{{\sqsubset}\xspace}
\newcommand{\mole}{{\sqsubseteq}\xspace}
\newcommand{\mogt}{{\sqsupset}\xspace}
\newcommand{\moge}{{\sqsupseteq}\xspace}
\newcommand{\statetransition}[3]{#1 \xrightarrow{#3} #2\xspace}

\newcommand{\subseq}[2]{#1 \subseteq_{o} #2\xspace}
\newcommand{\emptysequence}{\langle\rangle\xspace}
\newcommand{\hd}[1]{#1{:}hd\xspace}
\newcommand{\tl}[1]{#1{:}tl\xspace}
\newcommand{\pre}[2]{\texttt{prefix}_{[#1]}(#2)\xspace}
\newcommand{\lw}[2]{\texttt{LastW}_{[#1]}(#2)\xspace}
\newcommand{\pw}[2]{\texttt{prev}\mathbb{W}_{[#1]}(#2)\xspace}
\newcommand{\noup}[1]{{\nuparrow}{(#1)}\xspace}
\newcommand{\nodown}[1]{{\ndownarrow}{(#1)}\xspace}
\newcommand{\dep}[2]{\texttt{dep}(#1,#2)\xspace}

\newcommand{\eventreorder}[2]{#1 \xLeftarrow{R} #2\xspace}

\newcommand{\ob}[3]{#2 {<_{#1}} #3\xspace} 
\newcommand{\reln}[4]{#3 {\rightarrow^{#1}_{#2}} #4\xspace} 
\newcommand{\nreln}[4]{#3 \nrightarrow^{#1}_{#2} #4\xspace} 
\newcommand{\mo}[3]{\reln{\textbf{\textcolor{RedOrange}{mo}}}{#1}{#2}{#3}\xspace} 
\renewcommand{\to}[3]{\reln{\textbf{\textcolor{Brown}{to}}}{#1}{#2}{#3}\xspace} 
\newcommand{\csb}[3]{\reln{[c]{::}\textbf{\textcolor{Cerulean}{sb}}}{#1}{#2}{#3}\xspace} 
\newcommand{\crf}[3]{\reln{[c]{::}\textbf{\textcolor{PineGreen}{rf}}}{#1}{#2}{#3}\xspace} 
\newcommand{\cithb}[3]{\reln{[c]{::}\textbf{\textcolor{NavyBlue}{ithb}}}{#1}{#2}{#3}\xspace} 
\newcommand{\chb}[3]{\reln{[c]{::}\textbf{\textcolor{Maroon}{hb}}}{#1}{#2}{#3}\xspace} 
\newcommand{\nchb}[3]{\nreln{[c]{::}\textbf{\textcolor{Maroon}{hb}}}{#1}{#2}{#3}\xspace} 
\newcommand{\cmo}[3]{\reln{[c]{::}\textbf{\textcolor{RedOrange}{mo}}}{#1}{#2}{#3}\xspace} 
\newcommand{\cto}[3]{\reln{[c]{::}\textbf{\textcolor{Brown}{to}}}{#1}{#2}{#3}\xspace} 
\newcommand{\rs}[2]{\texttt{RS}_{#1}(#2)} 
\newcommand{\po}[3]{\reln{\textbf{\textcolor{CarnationPink}{po}}}{#1}{#2}{#3}\xspace} 
\newcommand{\rf}[3]{\reln{\textbf{\textcolor{PineGreen}{rf}}}{#1}{#2}{#3}\xspace} 
\newcommand{\dob}[3]{\reln{\textbf{\textcolor{Mulberry}{dob}}}{#1}{#2}{#3}\xspace} 
\newcommand{\sw}[3]{\reln{\textbf{\textcolor{Magenta}{sw}}}{#1}{#2}{#3}\xspace} 
\newcommand{\ithb}[3]{\reln{\textbf{\textcolor{NavyBlue}{ithb}}}{#1}{#2}{#3}\xspace} 
\newcommand{\ithbr}[3]{\reln{\textbf{\textcolor{NavyBlue}{ithb}-\textcolor{Lavender}{r}}}{#1}{#2}{#3}\xspace} 
\newcommand{\nithbr}[3]{\nreln{\textbf{\textcolor{NavyBlue}{ithb}-\textcolor{Lavender}{r}}}{#1}{#2}{#3}\xspace} 
\newcommand{\ithbi}[3]{\reln{\textbf{\textcolor{NavyBlue}{ithb}-\textcolor{Orange}{i}}}{#1}{#2}{#3}\xspace} 
\newcommand{\nithbi}[3]{\nreln{\textbf{\textcolor{NavyBlue}{ithb}-\textcolor{Orange}{i}}}{#1}{#2}{#3}\xspace} 
\newcommand{\hb}[3]{\reln{\textbf{\textcolor{Mahogany}{hb}}}{#1}{#2}{#3}\xspace} 
\newcommand{\nhb}[3]{\nreln{\textbf{\textcolor{Mahogany}{hb}}}{#1}{#2}{#3}\xspace} 
\newcommand{\mhb}[3]{#2 \textcolor{Red}{\twoheadrightarrow}_{#1} #3\xspace} 

\newcommand{\lcsb}{[c]{::}\textbf{\textcolor{Cerulean}{sb}}\xspace} 
\newcommand{\lcrf}{[c]{::}\textbf{\textcolor{PineGreen}{rf}}\xspace} 
\newcommand{\lcithb}{[c]{::}\textbf{\textcolor{NavyBlue}{ithb}}\xspace} 
\newcommand{\lchb}{[c]{::}\textbf{\textcolor{Maroon}{hb}}\xspace} 
\newcommand{\lcmo}{[c]{::}\textbf{\textcolor{RedOrange}{mo}}\xspace} 
\newcommand{\lcto}{[c]{::}\textbf{\textcolor{Brown}{to}}\xspace} 
\newcommand{\lpo}{\textbf{\textcolor{CarnationPink}{po}}\xspace} 
\newcommand{\lrf}{\textbf{\textcolor{PineGreen}{rf}}\xspace} 
\newcommand{\ldob}{\textbf{\textcolor{Mulberry}{dob}}\xspace} 
\newcommand{\lsw}{\textbf{\textcolor{Magenta}{sw}}\xspace} 
\newcommand{\lithb}{\textbf{\textcolor{NavyBlue}{ithb}}\xspace} 
\newcommand{\lhb}{\textbf{\textcolor{Mahogany}{hb}}\xspace} 
\newcommand{\lmo}{\textbf{\textcolor{RedOrange}{mo}}\xspace} 

\newcommand{\emcsb}{\csb{\tau}{}{}}

\newcommand{\emcithb}{\cithb{\tau}{}{}}
\newcommand{\emchb}{\chb{\tau}{}{}}
\newcommand{\emcmo}{\cmo{\tau}{}{}}
\newcommand{\emcto}{\cto{\tau}{}{}}
\newcommand{\empo}{\po{\tau}{}{}}
\newcommand{\emrf}{\rf{\tau}{}{}}
\newcommand{\emdob}{\dob{\tau}{}{}}
\newcommand{\emsw}{\sw{\tau}{}{}}
\newcommand{\emithb}{\ithb{\tau}{}{}}
\newcommand{\emithbr}{\ithbr{\tau}{}{}}
\newcommand{\emithbi}{\ithbi{\tau}{}{}}
\newcommand{\emhb}{\hb{\tau}{}{}}
\newcommand{\emmhb}{\mhb{\tau}{}{}\xspace} 
\newcommand{\emmo}{\mo{\tau}{}{}}
\newcommand{\emto}{\to{\tau}{}{}}

\newcommand{\tid}[1]{\textsf{tid}_{#1}}

\stackMath

\begin{document}
\title{{\fontsize{24pt}{1cm}\selectfont Dynamic Verification of C11 Concurrency over Multi Copy Atomics}}
%
%
%

\author{\IEEEauthorblockN{Sanjana Singh,
		Divyanjali Sharma and
		Subodh Sharma
	}
	\IEEEauthorblockA{Department of Computer Science $\&$ Engineering,
		Indian Institute of Technology Delhi\\
		Email: $\{$sanjana.singh,divyanjali,svs$\}$@cse.iitd.ac.in
	}
}

\maketitle              

\begin{abstract}
  We investigate the problem of runtime analysis of concurrent \cmodel programs
  under {\em Multi-Copy-Atomic} semantics (\mca).  Under \mca, one
  can analyze program outcomes solely through interleaving and
  reordering of thread events. As a result, obtaining intuitive 
  explanations of program outcomes becomes straightforward. Newer 
  versions of ARM (ARMv8 and later), Alpha, and Intel's x-86 support 
  \mca.
Our tests reveal that state-of-the-art dynamic verification techniques
that analyze program executions under the \cmodel memory model
report safety property violations that can be interpreted as false
alarms under \mca semantics.
 Sorting the true from false violations
puts an undesirable burden on the user.
  In this work, we provide a dynamic verification technique
  (\ourtechnique) to analyze 
\cmodel program executions
  which are permitted under the \mca model.
  We restrict \cmodel happens-before relation and propose coherence rules 
  to capture precisely those \cmodel program executions which are allowed 
  under the \mca model. \ourtechnique's exploration of the
  state-space is based on the \sota dynamic verification algorithm,
  \sdpor.
  Our experiments validate that \ourtechnique captures all coherent
  \cmodel program executions, and is precise for the \mca model.

\end{abstract}

\section{Introduction} \label{sec:intro}
The relaxed memory orderings over concurrent memory accesses 
introduced in
the C/C++ 2011 ISO standard (\cmodel)~\cite{C11-standard}
has been the object of
intense study in the past decade 
%
%
The axiomatic specification of \cmodel~\cite{batty2011mathematizing} 
defines relaxed program behaviors as relations over memory accesses 
along with constraints on stores that can possibly affect each
load. The semantics are known to be complex
and for a fragment of the standard -- {\em release-acquire} -- 
the state-reachability problem is shown 
to be undecidable~\cite{abdulla2019verification}. More notable is that
many of the feasible program behaviors under \cmodel semantics may
never manifest on the underlying  architectures.

Therefore, an important desideratum is to engineer analysis
tools/techniques to analyze \cmodel programs under restricted
hardware models with {\em precision}.
%
%
%
In this work, we investigate the problem of precise dynamic analysis
of \cmodel programs under {\em Multi-Copy-Atomic} model (\mca) -- a
popular hardware memory model, which is claimed to be supported in
Intel's x-86 TSO, newer versions of ARM\footnotemark (v8 and
later) and Alpha with varying degrees of permitted reorderings. A
noteworthy aspect of the \mca model is the assumption of a single abstract
view of shared memory between processing elements (or threads), leading
to the observation that permitted program behaviors under this model
can be explained solely through interleaving and reordering of thread
events.
As a result, one can use existing dynamic 
analyses~\cite{abdulla2014optimal}\cite{flanagan2005dynamic}\cite{abdulla2015stateless}\cite{zhang2015dynamic}
(with suitable adaptations)
developed for memory consistency models where program 
behaviors can be explained by program event interleavings alone.

\setlength{\skip\footins}{0.2cm}
\footnotetext{
	ARMv8 calls its model 
	{\em other}-\mca \cite{ARM-standard}. The difference with \mca is   
	of terminology, not semantics 
	(as clarified by \cite{pulte2017simplifying}).}
%
%



%

Several past works on designing efficient dynamic verifiers (also
referred to as stateless model checkers) for \cmodel concurrency 
(or its variants) exist such as 
\cds~\cite{norris2013cdschecker}
and GenMC~\cite{Raad19}.
On the one hand, notwithstanding the sophistication of their
algorithms, some of the valid \cmodel outcomes flagged by these tools, 
would not occur when the input program is executed on
an \mca architecture.
Consider the program (\hlref{IRIW}).
The initial value of $x$ and $y$ is $0$ and the value in brackets
(\ie $y(0)$ and $x(0)$) indicate the value read in local variables
$a$ and $b$.
No two statements in the program
can reorder due to the control dependence between the read statements
of threads $T_2$ and $T_4$. The execution shown in the program with $a {=} 0$
$\wedge$ $b{=}0$ is a valid \cmodel outcome if not all statements in the
program are \sc (sequential consistency) ordered.
It is so because the dependencies shown via gray dashed arrows may not
hold
under the \cmodel model, thus leading to an acyclic dependence relation
(explained further in Section~\refsection{sec:c11 hb}).
However, such an outcome cannot be explained through any {\em acyclic
	interleaving} of events; therefore, is invalid under the \mca
model. 
%
Hence,
if a property violation is detected by a \cmodel verification 
technique it must be scrutinized further for viability
under the given hardware memory model.
The laborious task of sorting the feasible from infeasible 
violations is a burden on the user.
On the other hand, solutions for \mca hardware memory models 
either ignore the \cmodel relaxation directives provided in the program such 
as~\cite{abdulla2015stateless}\cite{zhang2015dynamic} for x-86 TSO 
or have been designed for a memory model~\cite{podkopaev2019bridging} 
proposed as a superset of existing hardware models (including ARMv8), 
namely HMC~\cite{kokologiannakis2020hmc}.
\begin{figure}
	\centering
	\setlength{\tabcolsep}{0em}
	\begin{tabular}{m{0.85\columnwidth}m{0.10\columnwidth}}
		\tikzset{every picture/.style={line width=0.75pt}} 
\begin{tikzpicture}[x=1em,y=1em,yscale=1,xscale=1]
\tikzstyle{every node}=[font=\small]

\node (init) {$Initially$, $x=0$, $y=0$};
\node (wx1) [below left=15pt and 20pt of init] {$ x := 1 $};
\node (rx1) [right=20pt of wx1] {if$ (x=1) $};
\node (ry0) [below=8pt of rx1] {\tab$a := y(0) $};
\node (wy1) [right=20pt of rx1] {$ y := 1 $};
\node (ry1) [right=20pt of wy1] {if$ (y=1) $};
\node (rx0) [below=6pt of ry1] {\tab$b := x(0) $};
\draw [->,>=stealth,color=Maroon] (wx1) -- (rx1);
\draw [->,>=stealth,color=Maroon] (rx1)+(0,-3.5pt) -- (ry0);
\draw [->,>=stealth,color=gray, densely dashed] (ry0) -- (wy1);
\draw [->,>=stealth,color=Maroon] (wy1) -- (ry1);
\draw [->,>=stealth,color=Maroon] ($ (ry1.south)+(0,4pt) $) -- ($ (rx0.north)+(0,-3pt) $);
\draw [color=gray,densely dashed] (rx0.south)+(0,3pt) -- ($ (ry1.south)+(0,-24pt) $);
\draw [color=gray,densely dashed] ($ (ry1.south)+(0,-24pt) $) -- ($ (wx1.south)+(0,-25.5pt) $);
\draw [->,>=stealth,color=gray,densely dashed] ($ (wx1.south)+(0,-25.5pt) $) -- (wx1);

\node (T1) [above=-1pt of wx1] {$T_1$};
\node (T2) [above=-1pt of rx1] {$T_2$};
\node (T3) [above=-1pt of wy1] {$T_3$};
\node (T4) [above=-1pt of ry1] {$T_4$};

\end{tikzpicture}
		& (\hl{IRIW})	
	\end{tabular}
	\label{fig: iriw}
\end{figure}

Our contribution sits in this cross-section of \cmodel program
execution on \mca architectures.  We design a sound and precise
dynamic verification technique called \ourtechnique (pronounced as
moh{$\cdot$}kaa), addressing the problem of \cmodel program
verification under \mca.
The key contribution of our work is in restricting \cmodel program
behaviors to only those that are permitted under the \mca model. To accomplish 
this, we
present {\em happens-before} and {\em coherence} rules 
(see \refsection{sec:our hb}). Another central 
contribution of our work is to simulate 
the reordering of events of a thread (either by the compiler or the hardware)
through a new event type, \viz, {\em shadow-writes}. It 
updates the shared memory for the store instructions in the program.
%

Our proposed contributions have several merits: (M1) our
formalization makes \sdpor~\cite{abdulla2014optimal} 
(a powerful stateless model checking algorithm)
eminently usable without any modification; 
(M2) the use of shadow-writes simulates reordering through
interleaving and thus allows scheduling of execution sequences 
that reflect the effects
of reordering with simply interleaving of program events. 
%
%
%

The remaining paper is organized as follows: 
\refsection{sec:preliminaries} introduces assumptions and  
notations used;
\refsection{sec:mca} describes the segments of \mca operational semantics
proposed by Colvin and Smith~\cite{colvin2018wide}
that are relevant to this work;
\refsection{sec:c11 hb} briefly introduces the
\cmodel model \cite{C11-standard}\cite{batty2011mathematizing};
we present our proposed technique in \refsection{sec:our hb} and 
\refsection{sec:algo}, where
we formally introduce shadow-writes and present \ourtechnique's 
happens-before relation along with the coherence rules for 
soundness in \refsection{sec:our hb};
in \refsection{sec:algo} we explain the working of 
\ourtechnique technique; and,
present our experimental observations in 
\refsection{sec:results}.
To summarize, in this work we make the following contributions:
\begin{itemize}
	\item We introduce shadow-writes to precisely capture the
	reordering semantics of \mca through interleaving.
	
	\item We propose a restriction of \cmodel happens-before relation 
	to disallow non-\mca behaviors, and establish coherence
	rules to ensure exploration of only coherent \cmodel
	execution sequences.
	
	\item We establish the correctness and precision of our proposed 
	relations and rules with supporting theorems 
	(see Theorem~\ref{thm:coherence}, \refsection{sec:our hb} and 
	Theorem~\ref{thm:traces eq}, \refsection{sec:algo}).
	
	\item We present the \ourtechnique technique that executes \sdpor
	utilizing our proposed happens-before restriction 
	(see \refsection{sec:algo}).
	
	\item We present a prototype implementation to validate our
	technique and perform experiments to evaluate our
	claims empirically (see \refsection{sec:results}).
\end{itemize}

\section{Preliminaries} \label{sec:preliminaries}
\noindent {\bf Concurrent program model:}
We consider an acyclic multi-threaded program, $P$, as a finite set 
of program threads.
A thread $i$ in $P$, uniquely identified by $\tid{i}$, has 
deterministic computation and terminating executions.
%
The threads in $P$ access a fixed set of memory locations called objects
(denoted by $\objects$).
Each thread executes a sequence of {\em events} that,
in essence, are {\em actions} on these objects.
The set of actions (denoted by $\actions$) include: \textit{write},
\textit{read}, \textit{rmw} (or read-modify-write), 
and \textit{fence}. We extend this set with a special action
called \textit{shadow-write}. In our model of computation,
each \textit{write} or \textit{rmw} action is now associated with a
corresponding \textit{shadow-write} action.  This action updates the
shared memory with the value of its corresponding write or
rmw. 

Let $\states$ be the set of global states of $P$ with a given initial
state $s_0 \in \states$. We assume the standard definition of a
state, \ie, valuation of all shared and local objects of $P$, and the
program counter of all threads. In a global state
$\sigma \in \states$, {\bf shr} $\sigma$ denotes the shared component
comprising of (shared object, value) pairs, and {\bf lcl} $\sigma$
denotes the local component comprising of (local object, value) pairs.

%
%
%

\noindent {\bf Program execution and events.}  The set of all events of a
program $P$ is denoted by $\events$.  An execution of 
 $P$ is, therefore, a sequence of events $\tau {=} e_1. e_2. \ldots e_n$
s.t. $e_i \in \events$, for $i \in \{1, 2, \ldots, n\}$.
The sequence of events of a thread $\tid{i}$ is
denoted by $\tid{i}{:}\tau$.
The system upon execution of $e_i$
(which is a sequence of internal operations on local
objects followed by a single operation on a shared object)
 transitions from a state $s_{i-1}$ to the next state
$s_i$ (denoted by $\statetransition{s_{i-1}}{s_i}{e_i}$).
Note that the event $e_i$ must be {\em enabled} in state $s_{i-1}$ for
the transition to take place. 
%
Let $s_{[\tau]}$ represent the state reached after
exploring the execution sequence $\tau$ and $\pre{\tau}{e}$ represent the
prefix of $\tau$ up to (but not including) $e$.
Two events $e',e\in \events_\tau$ (where $\events_\tau$ 
denotes the set of events in $\tau$)
 are related by a total order
$\ob{\tau}{}{}$. For instance, $\ob{\tau}{e'}{e}$ denotes that $e'$ 
occurs before $e$ in $\tau$.
An empty sequence is represented as $\emptysequence$.
An event from thread $thr$ at index $idx$ in an execution $\tau$ is a tuple 
$\langle thr, act, obj, ord, idx \rangle$, where 
$act \in \mathcal{A}$ represents the action 
on a set of objects $obj \subseteq \mathcal{O}$ 
under the memory ordering constraint $ord \in \mathcal{M}$. 
Observe that $obj$ can potentially be a non-singleton for rmw events 
and empty set for fences. The projections $thr(e)$, $act(e)$, $obj(e)$, $ord(e)$, 
and $idx(e)$ return the respective tuple elements of $e$.
With the exception of $obj(e)$ when $e$ is an rmw event, all other
projections are straightforward to interpret.  When an rmw event $e$
is a read of $o_1$ and write of $o_2$, then $obj(e)$ returns $o_1$
when $e$ is analysed as a read event, and $o_2$ otherwise.

\noindent  {\bf Memory ordering constraints.}
Under the \cmodel model each event has an associated \textit{memory
order}. A memory order specifies the ordering possibilities of an (atomic or 
non-atomic) event around an atomic event; thereby restricting the freedom 
available to compilers and underlying systems to reorder events. 
Let $\memords$ = $\{$\na, \rlx, \rel, \acq, \acqrel, {\sc}$\}$ represent the 
set of memory orders -- relaxed (\rlx), release (\rel), acquire/consume (\acq), 
acquire-release (\acqrel) and 
sequentially-consistent (\sc) -- provided for atomic events and 
\na representing a non-atomic access.
Let $\molt {\subseteq} \memords {\times} \memords$ be a relation on memory 
orders such that $m_1 \molt m_2$ denotes that $m_2$ is stricter 
than $m_1$; thus, $m_2$ may restrict certain program outcomes
that are otherwise possible with $m_1$. 
The memory orders in $\memords$ are related as \na 
$\molt$ \rlx $\molt$ $\{$\texttt{acq,rel}$\}$ $\molt$ \acqrel 
$\molt$ \sc.
Accordingly, $\mole$ represents strict or stricter ordering. For instance,
\rel $\mole$ \rel.
We overload the operators $\molt$, $\mole$ as unary operators that
return the set of memory orders that are weaker. For instance,  
$\mole$\rel = $\{$\na, \rlx, \rel$\}$.
The set of events pertaining to a  memory order $m$ is represented
as $\ordevents{m}$. 
We also use $\ordevents{\molt m}$ (or $\ordevents{\mole m}$) to
represent the set of $\ordevents{m_1} \union
... \union \ordevents{m_N}$ where $m_i \molt m$ (or $m_i \mole m$), eg
$\ordevents{\mole\rlx}$ = $\ordevents{\na}$ $\union$
$\ordevents{rlx}$. Similarly, $\ordevents{\mogt m}$ (or
$\ordevents{\moge m}$) can be interpreted as union of sets of events
with ordering annotations stricter (or at least as strict) than (as) $m$.

\noindent {\bf A note on event categories.} For ease of explanation, we 
categorise the set of events in $\events$ into:  
(i)  set of writes ($\writes$)
that issue a write (\ie, write events and rmws), 
(ii) the set of modifiers ($\modifiers$) that update the shared memory for
an issued write (ie, shadow-write events), (iii) the set of
reads ($\reads$) (\ie, read events and rmws),
and (iv) the set of program fences ($\fences$).
Note that the shared-read events and shared-write events (including
corresponding shadow-writes) can either be atomic or non-atomic in
nature.
We use  $\writes_\tau, \reads_\tau, \modifiers_\tau$ and $\fences_\tau$ 
for the respective categories of the events in an execution $\tau$.
Similarly, we write $\ordwrites{m}_\tau, \ordreads{m}_\tau,
\ordmodifiers{m}_\tau$ and $\ordfences{m}_\tau$ to denote the set of events with 
associated memory order $m$ of the respective action categories in the execution 
sequence $\tau$.
%

\noindent {\bf Traces and equivalence relation:} 
Given $\tau$, if reversing the order of execution of
two co-enabled events $e',e \in \events_\tau$ does not change the outcome of 
the program then $e',e$ are called {\em independent} events \cite{mc-book}.
%
%
All  sequences that differ only in the order of independent events 
are called {\em equivalent} sequences. 
We use $\tau_1 \equi 
\tau_2$ to denote that $\tau_1$ and $\tau_2$ are equivalent.
A set of equivalent sequences form an equivalence class known as 
a program {\em trace} that represents a specific
behavior of the program.
A sound analysis must explore at least one sequence from each trace of $P$.

%

\section{Multi Copy Atomic model} \label{sec:mca}
Under the \mca model if a write to object $o$ from $\tid{i}$ is {\em
	observed} by a different thread, $\tid{j}$, 
then the write is {\em coherently} observable to
all other threads that access the object $o$. It is, however,
permitted for a thread to observe its own writes prior to making them
visible to other threads in the system.
%
The term {\em observed} refers to $\tid{j}$
becoming aware of a write from $\tid{i}$, either directly 
when a read of $\tid{j}$ reads-from the write or indirectly through a 
chain of intra- and inter-thread dependencies~\cite{ARM-standard}.
We formally revisit the term {\em observed} and  {\em coherent} access in
\refsection{sec:our hb} after formally defining event 
interactions.
%

We adapt the \mca model formalized by Colvin and
Smith~\cite{colvin2018wide} as a basis for our technique that 
relies on {\em shadow-writes}.
The model allows for a sequence of events 
$\tid{i}{:}\tau$ of thread $\tid{i}$ to be reordered to a sequence 
$\tid{i}{:}\tau'$.
%
Consider two events, $e', e \in  \events_{\tid{i}{:}\tau}$.
Assume that $e$ originally occurred {\em after} $e'$. 
We represent the reordering such that $e$ now occurs {\em before} $e'$ 
by $\eventreorder{e'}{e}$.
The reordering $\eventreorder{e'}{e}$ can take place only 
if $e',e$ have distinct local and shared variables.
To ensure a semantic preserving reordering, 
the following must hold:
\begin{itemize}[\IEEEsetlabelwidth{spr1}]
	\item  [spr1] $\events_{\tid{i}{:}\tau'}
	= \events_{\tid{i}{:}\tau}$ (the event sets are the same)
	\item [spr2] each
	read event of $\tid{i}{:}\tau$ must have the same set of writes to
	read from in $\tid{i}{:}\tau$ as well as $\tid{i}{:}\tau'$ \newline
	(we call it {\em thread  semantics})
	\item [spr3] $\tid{i}{:}\tau'$ preserves  
	the order of updates and accesses of each shared variable 
	with respect to $\tid{i}{:}\tau$ \newline
	(we call it {\em coherence-per-location}).
\end{itemize}
%
%
%
%
%

\noindent
The following language represents an \mca
model \cite{colvin2018wide}.

\ \\
\noindent\fbox{\begin{minipage}[t]{0.47\textwidth}
		\small
		\noindent
		processing element, $p := ({\bf tid}_N\ {\bf lcl}\ \sigma 
		\boldsymbol{\cdot} \tid{N}{:}\tau) \| p1 \parallel p2$
		
		\noindent
		system,\tab\tab\quad $\ s := ({\bf shr}\ \sigma \boldsymbol{\cdot} p)$
\end{minipage}}
\ \\ \ \\
\noindent
The key element of the language is a {\em processing element}. 
It is identified by a unique identifier ({\bf tid$_N$}), the 
local state ({\bf lcl} $\sigma$), 
and a sequence of events to 
be executed ($\tid{N}{:}\tau$). The entity {\em system} 
is identified with a shared state ({\bf shr} 
$\sigma$) and a parallel 
composition of processing elements.
Thus, a system term is of the form 
{\small
	$({\bf shr}\ \sigma \boldsymbol{\cdot} 
	({\bf tid}_0\ {\bf lcl}\ \sigma \boldsymbol{\cdot} \tid{0}{:}\tau_0) \parallel
	({\bf tid}_1\ {\bf lcl}\ \sigma \boldsymbol{\cdot} \tid{1}{:}\tau_1) \parallel
	\ldots)$}.

	A dynamic verification technique executes the input program $P$. At
	each state of the execution, the verifier ascertains the set
	containing the next event from each processing element to be
	explored.
	Thus the program for each thread is assumed to be  
	a linear sequence of events without branches.
%
%
The operational semantics for a program \program under \mca is
listed in \reffig{fig: mca semantics}.

\begin{figure}
	\def\arraystretch{2.5}
	\begin{center}
		\begin{tabular}{|cc|}
			\hline
			$\frac{\statetransition{\tid{i}{:}e.\tau'\ }{\ \tid{i}{:}\tau}{e} \quad \eventreorder{e'\ }{\ e}}{\statetransition{\tid{i}{:}e'.e.\tau'\ }{\ \tid{i}{:}e'.\tau}{e}}$&{\footnotesize \hl{(reordering)}} 
			\\
			
			$\frac{\statetransition{p_1\ }{\ p_1'}{e}}{\statetransition{p_1 \parallel p_2\ }{\ p_1' \parallel p_2}{e}}$
			$\frac{\statetransition{p_2\ }{\ p_2'}{e}}{\statetransition{p_1 \parallel p_2\ }{\ p_1 \parallel p_2'}{e}}$ &{\footnotesize\hl{(parcom)}}
			\\
			
			$\frac{\statetransition{\tau'\ }{\ \tau}{r:=x} \quad {\bf shr}\ \sigma(x)\ =\ v}{\statetransition{({\bf lcl}\ \sigma \boldsymbol{\cdot} \tau')\ }{\ ({\bf lcl}\ \sigma_{[r:=v]} \boldsymbol{\cdot} \tau)}{[x=v]}}$ &{\footnotesize \hl{(r-shared)}}
			\\
			
			$\frac{\statetransition{\tau'\ }{\ \tau}{x := r} \quad \sigma(r)\ =\ v}{\statetransition{({\bf lcl}\ \sigma \boldsymbol{\cdot} \tau')\ }{\ ({\bf lcl}\ \sigma \boldsymbol{\cdot} \tau)}{x := v}}$ &{\footnotesize \hl{(w-issue)}}
			\\
			
			$\frac{\statetransition{p\ }{\ p'}{\tid{i}::x:=v}}
			{\quad \statetransition{({\bf shr}\ \sigma \boldsymbol{\cdot}\ p)\ }{\ ({\bf shr}\ \sigma_{[x:=v]} \boldsymbol{\cdot}\ p')}{*}}$ &{\footnotesize \hl{(w-update)}}
			\\
			\hline
	\end{tabular} \end{center}
	\caption{Semantics of \mca model~\cite{colvin2018wide}}
	\label{fig: mca semantics}
\end{figure}

\noindent {\bf \mca semantics.}
The {\hlref{(reordering)}} rule states that if an event $e$ can 
reorder before another event $e'$ (where $\ob{\tid{i}{:}\tau}{e'}{e}$) 
then the processing element $\tid{i}$ can execute $e$ before $e'$ 
and suitably update the remaining sequence to be executed later.
%
The rule \hlref{(parcom)} shows the parallel composition of the 
processing elements. It states that one step of the system is taken
by one processing element at a time.
The \hlref{(r-shared)} rule captures the read of a shared
variable from the shared storage into a local variable. 
The \hlref{(w-issue)} rule shows that a processing element 
initiates a write operation of value $v$ to a shared variable $x$, 
and moves to the next event.
%
A write initiated by $\tid{i}$ is updated
to the shared storage by the system 
as shown in rule \hlref{(w-update)}. Notably, this rule
captures the effects of {\em shadow-write}
events.

\section{\cmodel Memory Model Semantics} \label{sec:c11 hb}
\begin{figure*}[t!]
	\begin{minipage}{\textwidth}
		\centering
		\resizebox{0.995\textwidth}{!}{
			\begin{tabular}{|c|c|c|c|c|c|}
				\hline
				\resizebox{0.15\textwidth}{!}{\tikzset{every picture/.style={line width=0.75pt}} 
\begin{tikzpicture}[x=1em,y=1em,yscale=-1,xscale=-1]
\tikzstyle{every node}=[font=\normalfont]
\node (ew1) {$ e_{w1} $};
\node (ew2) [below=20pt of ew1] {$ e_{w2} $};
\node (mo1) [below=0pt of ew2] {(mo1)};

\draw [->,>=stealth,color=RedOrange,thin] ($ (ew1.south east)+(.5,-5pt) $) to[out=135,in=-135] node[midway,right=-2pt,font=\scriptsize] {\textcolor{black}{\lcmo}} ($ (ew2.south east)+(0.4,-5pt) $);
\draw [->,>=stealth,color=Mahogany,thin] ($ (ew1.south west)+(-.3,-5pt) $) to[out=45,in=-45] node[midway,left=-2pt,font=\scriptsize] {\textcolor{black}{\lchb}} ($ (ew2.south west)+(-0.3,-5pt) $);

\end{tikzpicture}} &
				\resizebox{0.15\textwidth}{!}{\tikzset{every picture/.style={line width=0.75pt}} 
\begin{tikzpicture}[x=1em,y=1em,yscale=-1,xscale=-1]
\tikzstyle{every node}=[font=\normalfont]
\node (er1) [inner sep=1pt] {$ e_{r1} $};
\node (er2) [right=25pt of er1, inner sep=1pt] {$ e_{r2} $};
\node (ew1) [below left=20pt and -15pt of er1, inner sep=1pt] {$ e_{w1} $};
\node (ew2) [below left=20pt and -15pt of er2, inner sep=1pt] {$ e_{w2} $};
\node (mo2) [below right=5pt and -1pt of ew1, inner sep=1pt] {(mo2)};

`\draw [->,>=stealth,color=Mahogany,thin] (er1) -- node[midway,above=-2pt,font=\scriptsize,color=black] { $\lchb$ } (er2);
\draw [->,>=stealth,color=RedOrange,thin] (ew1) -- node[midway,below=-2pt,font=\scriptsize,color=black] { $ \lcmo $ } (ew2);
\draw [->,>=stealth,color=PineGreen,thin] (ew1) -- node[midway,left=-2pt,font=\scriptsize,color=black] { $\lcrf$ } (er1);
\draw [->,>=stealth,color=PineGreen,thin] (ew2) -- node[midway,left=-2pt,font=\scriptsize,color=black] { $\lcrf$ } (er2);


\end{tikzpicture}} &
				\resizebox{0.15\textwidth}{!}{\tikzset{every picture/.style={line width=0.75pt}} 
\begin{tikzpicture}[x=1em,y=1em,yscale=-1,xscale=-1]
\tikzstyle{every node}=[font=\normalfont]
\node (er1) [inner sep=1pt] {$ e_{r1} $};
\node (ew1) [right=25pt of er1, inner sep=1pt] {$ e_{w1} $};
\node (ew2) [below left=20pt and -15pt of er1,inner sep=1pt] {$ e_{w2} $};
\node (mo3) [below right=0pt and 8pt of ew2, inner sep=1pt] {(mo3)};

\draw [->,>=stealth,color=Mahogany,thin] (er1) -- node[midway,above=-2pt,font=\scriptsize,color=black] { $\lchb$ } (ew1);
\draw [->,>=stealth,color=RedOrange,thin] (ew2) -- node[midway,right=2pt,font=\scriptsize,color=black] { $\lcmo$ } (ew1);
\draw [->,>=stealth,color=PineGreen,thin] (ew2) -- node[midway,left=-2pt,font=\scriptsize,color=black] { $\lcrf$ } (er1);`


\end{tikzpicture}} &
				\resizebox{0.15\textwidth}{!}{\tikzset{every picture/.style={line width=0.75pt}} 
\begin{tikzpicture}[x=1em,y=1em,yscale=-1,xscale=-1]
\tikzstyle{every node}=[font=\normalfont]
\node (ew1) [inner sep=1pt] {$ e_{w1} $};
\node (er1) [right=25pt of ew1, inner sep=1pt] {$ e_{r1} $};
\node (ew2) [below left=20pt and -15pt of ew1,inner sep=1pt] {$ e_{w2} $};
\node (mo4) [below right=0pt and 8pt of ew2, inner sep=1pt] {(mo4)};

\draw [->,>=stealth,color=Mahogany,thin] (ew1) -- node[midway,above=-2pt,font=\scriptsize,color=black] { $\lchb$ } (er1);
\draw [->,>=stealth,color=PineGreen,thin] (ew2) -- node[midway,right=2pt,font=\scriptsize,color=black] { $\lcrf$ } (er1);
\draw [->,>=stealth,color=RedOrange,thin] (ew1) -- node[midway,left=-2pt,font=\scriptsize,color=black] { $\lcmo$ } (ew2);`


\end{tikzpicture}} &
				\resizebox{0.15\textwidth}{!}{\tikzset{every picture/.style={line width=0.75pt}} 
\begin{tikzpicture}[x=1em,y=1em,yscale=-1,xscale=-1]
\tikzstyle{every node}=[font=\normalfont]
\node (esc1) {$ e_{sc1} $};
\node (esc2) [below=20pt of esc1] {$ e_{sc2} $};
\node (to) [below=0pt of esc2] {(to)};

\draw [->,>=stealth,color=RedOrange,thin] ($ (esc2.south east)+(.3,-5pt) $) to[out=-135,in=135] node[strike out,draw,above=2pt,color=black,-]{} node[right=-2pt,pos=.5,font=\scriptsize] {\textcolor{black}{\lcmo/}} node[right=-0.5pt,pos=.25,font=\scriptsize] {\textcolor{black}{\lchb}} ($ (esc1.south east)+(0.4,-5pt) $);
\draw [->,>=stealth,color=Brown,thin] ($ (esc1.south west)+(-.3,-5pt) $) to[out=45,in=-45] node[midway,left=-2pt,font=\scriptsize] {\textcolor{black}{\lcto}} ($ (esc2.south west)+(-0.3,-5pt) $);

\end{tikzpicture}} &
				\resizebox{0.15\textwidth}{!}{\tikzset{every picture/.style={line width=0.75pt}} 
\begin{tikzpicture}[x=1em,y=1em,yscale=-1,xscale=-1]
\tikzstyle{every node}=[font=\normalfont]
\node (ewx) {$ e_{wx} $};
\node (erx) [below=20pt of ewx] {$ e_{rx} $};
\node (co) [below=0pt of erx] {(co)};

\draw [->,>=stealth,color=Mahogany,thin] ($ (erx.south east)+(.3,-5pt) $) to[out=-135,in=135] node[strike out,draw,above=2pt,color=black,-]{} node[midway,right=-2pt,font=\scriptsize] {\textcolor{black}{\lchb}} ($ (ewx.south east)+(0.4,-5pt) $);
\draw [->,>=stealth,color=PineGreen,thin] ($ (ewx.south west)+(-.3,-5pt) $) to[out=45,in=-45] node[midway,left=-2pt,font=\scriptsize] {\textcolor{black}{\lcrf}} ($ (erx.south west)+(-0.3,-5pt) $);
\end{tikzpicture}} \\
				\hline
		\end{tabular}}
		\newline
		\resizebox{0.995\textwidth}{!}{
			\begin{tabular}{|p{0.04\textwidth} p{0.95\textwidth}|}
				(\hl{mo1}) & 	$\forall e_{w1},e_{w2} \in \writes_\tau$, 
				$\chb{\tau}{e_{w1}}{e_{w2}} \implies \cmo{\tau}{e_{w1}}{e_{w2}}$ 
				\\
				
				(\hl{mo2}) & $\forall e_{r1},e_{r2} {\in} \reads_\tau,$ $e_{w1} {\in} 
				\writes_\tau,$ $\chb{\tau}{e_{r1}}{e_{r2}}$ $\^$
				$\crf{\tau}{e_{w1}}{e_{r1}}$ 
				$\implies$ $\crf{\tau}{e_{w1}}{e_{r2}}$ 
				$\v$ $(\exists e_{w2},$ $\crf{\tau}{e_{w2}}{e_{r2}}$ $\^$ 
				$\cmo{\tau}{e_{w1}}{e_{w2}})$ 
				\\
				
				(\hl{mo3}) & $\forall e_{r1} \in \reads_\tau,\ e_{w1} \in \writes_\tau,\ 
				\chb{\tau}{e_{r1}}{e_{w1}} \implies \exists e_{w2}, 
				\cmo{\tau}{e_{w2}}{e_{w1}} \st \crf{\tau}{e_{w2}}{e_{r1}}$
				\\
				
				(\hl{mo4}) & $\forall e_{w1} \in \writes_\tau,\ e_{r1} \in \reads_\tau,
				\chb{\tau}{e_{w1}}{e_{r1}} \implies \crf{\tau}{e_{w1}}{e_{r1}} \v
				(\exists e_{w2},\ \crf{\tau}{e_{w2}}{e_{r1}} \^ 
				\cmo{\tau}{e_{w1}}{e_{w2}})$ 
				\\
				
				(\hl{to}) & $\nexists e_{sc1},e_{sc2} \in \events_\tau^{(\sc)}$ $\st$ 
				$\cto{\tau}{e_{sc1}}{e_{sc2}}$ $\^$ $(\chb{\tau}{e_{sc2}}{e_{sc1}}$ $\v$ 
				$\cmo{\tau}{e_{sc2}}{e_{sc1}})$ 
				\\
				
				(\hl{co}) & $\forall e_r \in \reads_\tau$, $\exists e_w \in \writes_\tau$ 
				\st $\crf{\tau}{e_w}{e_r}$ $\^$ $\nchb{\tau}{e_r}{e_w}$ 
				\\\hline
				
		\end{tabular}}
		\caption{\cmodel coherence rules}
		\label{fig: c11 rules}
	\end{minipage}
	\vspace{-1em}
\end{figure*}

In \cmodel memory model the behavior of an execution, 
$\tau$, is usually defined through an acyclic and irreflexive 
happens-before relation 
($\emchb \subseteq  \events_{\tau} \times \events_{\tau}$). 
%
The \cmodel model defines its happens-before relation $\chb{}{}{}$ =
$\emcsb \union \emcithb$, 
\setlength{\textfloatsep}{0pt}
where: (i) $\emcsb$
({\it Sequenced-before}) is the intra-thread order on events, and (ii)
$\emcithb$ ({\it Inter-thread hb}) is the relation between
events of different threads (say $\tid{i}$, $\tid{j}$) formed by a
transitive closure of $\emcsb \union$ synchronizations (that occur when a 
read of $\tid{i}$ with an ordering $\moge \acq$ 
reads from a write event in a {\em release sequence}
of some event $e$ in $\tid{j}$) \cite{C11-standard}.

In addition, all write events of an object in a sequence $\tau$ must 
be related by a
total order called {\em modification-order} 
($\emcmo \subseteq \writes_\tau \times \writes_\tau$). 
The $\emcmo$ relation
is fundamentally involved in specifying
a set of sufficient conditions which ensure that 
a \cmodel program execution is {\em coherent}. 
The conditions are formally presented in \reffig{fig: c11 rules} along with
	their diagrammatic representations.
The conditions are as follows: 
\begin{itemize}
	\item (\hlref{mo1}): hb-ordered writes are also mo-ordered;
	\item (\hlref{mo2}): a read $e_{r2}$ hb-ordered after another read $e_{r1}$, 
	either reads-from the same source as $e_{r1}$'s or from a source 
	mo-ordered after the source of $e_{r1}$;
	\item (\hlref{mo3}): a read hb-ordered before a write reads-from a write 
	mo-ordered before that write;
	\item (\hlref{mo4}): a read hb-ordered after a write either reads-from the 
	write or from a write mo-ordered after that write;
	\item (\hlref{to}): all \sc ordered events must form a total-order 
	($\emcto$) \wrt $\emcmo$ and $\emchb$; and,
	\item (\hlref{co}): a read must take its data from a 
	write event occurring in the trace.
	The (\hlref{co}) rule ensures that a read does not take a value from 
	thin-air \ie a value not generated in the program execution.
\end{itemize}

Notice that shadow-write events
are a particular constuct of our technique; naturally, definitions of \cmodel
relations do not contain them. In \refsection{sec:our hb}, we shall present 
one of our main contributions of redefining the above-mentioned relations
(keeping shadow-write events in consideration) which admit only the 
\mca behaviors of a \cmodel program.



\noindent {\bf Reordering restrictions.}
For any two events $e',e \in \events_{\tau} \st 
thr(e) = thr(e') \wedge \ob{\tau}{e'}{e}$, we have the 
following: 
if $e \in \ordwrites{\moge \rel}$, then it restricts events such as $e'$
from reordering after it.
We denote this downward reordering restriction on $e'$ by $e$ as $\nodown{e,e'}$.
Similarly, if $e' \in \ordreads{\moge \acq}$, it restricts a {\em later} event
$e$ from reordering before it. 
We denote this upward reordering restriction on $e$ by $e'$ as $\noup{e', e}$.
Furthermore, \cmodel also disallows reordering of events $e',e$ that share program
dependence (such as data, address and control), which we represent by
$\dep{e'}{e}$.

\section{\mca restriction for \cmodel } \label{sec:our hb}
\begin{figure*}[t]
	
	\setlength{\tabcolsep}{2pt}
	\begin{tabular}{cc}
		\fbox{
			\begin{tabular}[c]{c}		
				\begin{minipage}{0.4\textwidth}
					\centering
					\setlength{\tabcolsep}{2pt}
					\begin{tabular}{p{0.2\linewidth}p{0.7\linewidth}}
						$\po{\tau}{e'}{e}$ & if
						$\csb{\tau}{e'}{e} \v act(e') = act(e) =$ shadow-write
						$\^ thr(e') = thr(e) \^ idx(e') < idx(e)$ \\\hline
						
						
						$\dob{\tau}{e'}{e}$ & if 
						$e' \in \ordwrites{\moge\rel}_\tau, e \in \ordreads{\moge\acq}_\tau 
						\^ \exists e'' \in \writes_\tau \st \rf{\tau}{e''}{e} \^ 
						e'' \in \events_{\rs{\tau}{e'}}$ \\\hline
						
						$\sw{\tau}{e'}{e}$ & if 
						$e' \in \ordwrites{\moge\rel}_\tau, e \in \ordreads{\moge\acq}_\tau 
						\^ \rf{\tau}{e'}{e}$ \\
						%
					\end{tabular}
				\end{minipage} 
				
				\vrule
				
				\begin{minipage}{0.36\textwidth}
					\centering
					\setlength{\tabcolsep}{2pt}
					\begin{tabular}[c]{p{0.2\linewidth}p{0.7\linewidth}}
						$\ithb{\tau}{e'}{e}$ & if
						$\sw{\tau}{e'}{e} \v \dob{\tau}{e'}{e} \v
						\exists e'' \st 
						(\sw{\tau}{e'}{e''} \^ \po{\tau}{e''}{e}) \v
						(\po{\tau}{e'}{e''} \^ \ithb{\tau}{e''}{e}) \v
						(\ithb{\tau}{e'}{e''} \^ \ithb{\tau}{e''}{e})$
						\newline
						\\\hline
						
						$\hb{\tau}{e'}{e}$ & if
						$\po{\tau}{e'}{e} \v \ithb{\tau}{e'}{e}$ \\\hline
						
						$\ \mhb{\tau}{e'}{e}$ & if
						$\hb{\tau}{e'}{e} \^ \neg (\sw{\tau}{e'}{e} 
						\v \dob{\tau}{e'}{e})$  \\
						
					\end{tabular}
				\end{minipage}
				
				%
				%
				%
			\end{tabular}	
		}\vspace{-1em}
		
		&
		
		\begin{minipage}{0.2\textwidth}
			\resizebox{\textwidth}{!}{\tikzset{every picture/.style={line width=0.75pt}} 
\begin{tikzpicture}[x=1em,y=1em,yscale=-1,xscale=-1]
\tikzstyle{every node}=[font=\normalfont]
\node (ewx1) {$ e^{\rel}_{wx1} $};
\node (erx1) [right=25pt of ewx1] {$ e^{\acq}_{rx} $};
\node (ewx2) [below left=20pt and -15pt of ewx1] {$ e_{wx2} $};
\node (i) [left=-2pt of ewx1] {(i)};

\draw [->,>=stealth,color=CarnationPink,thin] (ewx1) -- node[midway,left=-2pt,font=\scriptsize,color=black] { $\lpo$ } (ewx2);
\draw [->,>=stealth,color=Mulberry,thin] (ewx1) -- node[midway,above=-2pt,font=\scriptsize,color=black] { $\ldob$ } (erx1);
\draw [->,>=stealth,color=NavyBlue,thin] ($ (ewx1.east)+(0,.5pt) $) -- node[midway,below=-2pt,font=\scriptsize,color=black] { $\lithb$ } ($ (erx1.west)+(0,.5pt) $);
\draw [->,>=stealth,color=PineGreen,thin] (ewx2) -- node[midway,right,font=\scriptsize,color=black] { $\lrf$ } (erx1);

\draw [very thin, color=Gray] (-2,-1) -- (-2,4.7);
\draw [very thin, color=Gray] (-2.1,-1) -- (-2.1,4.7);


\end{tikzpicture}}
			\newline
			\resizebox{\textwidth}{!}{\tikzset{every picture/.style={line width=0.75pt}} 
\begin{tikzpicture}[x=1em,y=1em,yscale=-1,xscale=-1]
\tikzstyle{every node}=[font=\normalfont]
\node (ewx) {$ e^{\rel}_{wx} $};
\node (erx) [right=25pt of ewx] {$ e^{\acq}_{rx} $};
\node (ewy) [below left=20pt and -15pt of ewx] {$ e_{wy} $};
\node (ery) [below left=20pt and -15pt of erx] {$ e_{ry} $};
\node (ii) [left=-3pt of ewx] {(ii)};

\draw [->,>=stealth,color=Mulberry,thin] (ewx) -- node[above=-2pt,pos=.8,font=\scriptsize,color=black] { $\lsw$ } (erx);
\draw [->,>=stealth,color=PineGreen,thin] ($ (ewx.east)+(0,-1pt) $) -- node[above=-2pt,pos=.2,font=\scriptsize,color=black] { $\lrf$ } ($ (erx.west)+(0,-1pt) $);
\draw [->,>=stealth,color=NavyBlue,thin] ($ (ewx.east)+(0,1pt) $) -- node[pos=0.2,below=-2pt,font=\scriptsize,color=black] { $\lithb$ } ($ (erx.west)+(0,1pt) $);

\draw [->,>=stealth,color=CarnationPink,thin] (ewy) -- node[midway,left=-2pt,font=\scriptsize,color=black] { $\lpo$ } (ewx);
\draw [->,>=stealth,color=CarnationPink,thin] (erx) -- node[midway,left=-2pt,font=\scriptsize,color=black] { $\lpo$ } (ery);
\draw [->,>=stealth,color=NavyBlue,thin] (ewy) -- node[pos=0.3,above=-2pt,font=\scriptsize,color=black] { $\lithb$ } (ery);

\draw [very thin, color=Gray] (-2.5,-1) -- (-2.5,4.7);
\draw [very thin, color=Gray] (-2.6,-1) -- (-2.6,4.7);

\end{tikzpicture}}
		\end{minipage}
		
		%
	\end{tabular} 
	
	\caption{$\emhb$ relation}
	\label{fig:our hb}
	\vspace{-1em}
\end{figure*}

We (i) re-formulate \cmodel's $\emchb$ relation
and (ii) define trace coherence rules
to accurately recognize \cmodel
program traces admissible under \mca. 
A principal contribution of our technique is 
the introduction of a new event type called {\em shadow-writes}
that simulate reordering through interleaving
as explained below.
Shadow-writes
break a write operation into two (not necessarily consecutive)
events: (i) the write event from the program that is visible only to
events of the same thread and (ii) the shadow-write event that updates
the shared memory with the write event's value at a later timestamp;
thus, completing the write operation and making it visible to all
threads.
%
%
%
\noindent
In order to issue shadow-write events, we introduce 
the notion of  {\em shadow-threads}. 
We maintain a separate shadow-thread per program thread per object. 
%
For an event $e \in \writes$,
we use $shw(e)$ to represent the shadow-write event associated with 
$e$. 
Similarly, $prw(e')$ denotes the write event corresponding 
to the shadow-write $e' \in \modifiers$.
The set of shadow-threads associated with $thr(e)$ is denoted by 
$sth(e)$.
Note that shadow-writes of the threads in $sth(e)$ can interleave with 
the events of $thr(e)$ thereby enabling reordering through interleaving.

\begin{figure}[H]
	\centering
	\begin{tabular}{@{}c@{}c@{}}
		& \resizebox{0.65\columnwidth}{!}{\tikzset{every picture/.style={line width=0.75pt}} 
\begin{tikzpicture}[framed,x=1em,y=1em,yscale=-1,xscale=-1]
\tikzstyle{every node}=[font=\normalfont]
\node (i) {(i)};
\node (shared) [below left=-2pt and 0pt of i, rotate=90,font=\scriptsize] {shared};
\node (memory) [below right=27pt and -4pt of shared, rotate=90,font=\scriptsize] {memory};

\node (tau) [right=1pt of i,font=\large] {$ \tau = $};
\node (a) [right=-4pt of tau, color=blue] {(a)};
\node (alt) [right=-4pt of a] {$ <_{\tau} $};
\node (ap) [right=-4pt of alt, color=Purple] {(a')};
\node (aplt) [right=-4pt of ap] {$ <_{\tau} $};
\node (b) [right=-4pt of aplt, color=Magenta] {(b=1)};
\node (blt) [right=-4pt of b] {$ <_{\tau} $};
\node (c) [right=-4pt of blt, color=blue] {(c)};
\node (clt) [right=-4pt of c] {$ <_{\tau} $};
\node (d) [right=-4pt of clt, color=Magenta] {(d=2)};
\node (dlt) [right=-4pt of d] {$ <_{\tau} $};
\node (cp) [right=-4pt of dlt, color=Purple] {(c')};

\node (initx) [below left=0pt and -12pt of tau, font=\scriptsize] {$ x $};
\node (initval) [below=0pt of initx, font=\scriptsize] {0};
\begin{scope}[on background layer]
\node (initmem) [rectangle, draw, fit=(initx)(initval), fill=Gray!20, inner sep=-2pt] {};
\end{scope}

\node (ax) [below =-1.5pt of a, font=\scriptsize] {$ x $};
\node (aval) [below=0pt of ax, font=\scriptsize] {0};
\begin{scope}[on background layer]
\node (amem) [rectangle, draw, fit=(ax)(aval), fill=Gray!20, inner sep=-2pt] {};
\end{scope}

\node (apx) [below =-1.5pt of ap, font=\scriptsize] {$ x $};
\node (apval) [below=0pt of apx, font=\scriptsize] {1};
\begin{scope}[on background layer]
\node (apmem) [rectangle, draw, fit=(apx)(apval), fill=Gray!20, inner sep=-2pt] {};
\end{scope}

\node (bx) [below =-1.5pt of b, font=\scriptsize] {$ x $};
\node (bval) [below=0pt of bx, font=\scriptsize] {1};
\begin{scope}[on background layer]
\node (bmem) [rectangle, draw, fit=(bx)(bval), fill=Gray!20, inner sep=-2pt] {};
\end{scope}

\node (cx) [below =-1.5pt of c, font=\scriptsize] {$ x $};
\node (cval) [below=0pt of cx, font=\scriptsize] {1};
\begin{scope}[on background layer]
\node (cmem) [rectangle, draw, fit=(cx)(cval), fill=Gray!20, inner sep=-2pt] {};
\end{scope}

\node (dx) [below =-1.5pt of d, font=\scriptsize] {$ x $};
\node (dval) [below=0pt of dx, font=\scriptsize] {1};
\begin{scope}[on background layer]
\node (dmem) [rectangle, draw, fit=(dx)(dval), fill=Gray!20, inner sep=-2pt] {};
\end{scope}

\node (cpx) [below =-1.5pt of cp, font=\scriptsize] {$ x $};
\node (cpval) [below=0pt of cpx, font=\scriptsize] {2};
\begin{scope}[on background layer]
\node (cpmem) [rectangle, draw, fit=(cpx)(cpval), fill=Gray!20, inner sep=-2pt] {};
\end{scope}

\end{tikzpicture}}
		\\
		\multirow[t]{2}{*}{\resizebox{0.35\columnwidth}{!}{\tikzset{every picture/.style={line width=0.75pt}} 
\begin{tikzpicture}[x=1em,y=1em,yscale=-1,xscale=-1]
\tikzstyle{every node}=[font=\normalfont]
\node (t1) [] {$T_1$};
\node (t2) [right=35pt of t1] {$T_2$};
\node (a) [below=1pt of t1] {\textcolor{blue}{(a)} $x:=1$};
\node (b) [below=1pt of t2] {\textcolor{Magenta}{(b)} $l_1:=x$};
\node (c) [below=-2pt of b] {\textcolor{blue}{(c)} $x:=2$};
\node (d) [below=-2pt of c] {\textcolor{Magenta}{(d)} $l_2:=x$};

\node (st1) [below=33pt of a] {$sth_x(T_1)$};
\node (st2) [right=15pt of st1] {$sth_x(T_2)$};
\node (a1) [below=1pt of st1] {\textcolor{Purple}{(a')} $shw$\textcolor{blue}{(a)}};
\node (c1) [below=1pt of st2] {\textcolor{Purple}{(c')} $shw$\textcolor{blue}{(c)}};

\draw [very thin, color=Gray] (-2.6,-0.5) -- (-2.6,4.9);
\draw [very thin, color=Gray] (-2.7,-0.5) -- (-2.7,4.9);

\draw [very thin, color=Gray] (1.5,5.3) -- (-7,5.3);
\draw [very thin, color=Gray] (1.5,5.4) -- (-7,5.4);

\draw [very thin, color=Gray] (-2.6,6.0) -- (-2.6,9.0);
\draw [very thin, color=Gray] (-2.7,6.0) -- (-2.7,9.0);

\end{tikzpicture}}}& \resizebox{0.65\columnwidth}{!}{\tikzset{every picture/.style={line width=0.75pt}} 
\begin{tikzpicture}[framed,x=1em,y=1em,yscale=-1,xscale=-1]
\tikzstyle{every node}=[font=\normalfont]
\node (ii) {(ii)};

\node (tau) [right=0pt of ii,font=\large] {$ \tau = $};
\node (a) [right=-4pt of tau, color=blue] {(a)};
\node (alt) [right=-4pt of a] {$ <_{\tau} $};
\node (ap) [right=-4pt of alt, color=Purple] {(a')};
\node (aplt) [right=-4pt of ap] {$ <_{\tau} $};
\node (b) [right=-4pt of aplt, color=Magenta] {(b=1)};
\node (blt) [right=-4pt of b] {$ <_{\tau} $};
\node (c) [right=-4pt of blt, color=blue] {(c)};
\node (clt) [right=-4pt of c] {$ <_{\tau} $};
\node (d) [right=-4pt of clt, color=Magenta] {(d=2)};
\node (dlt) [right=-4pt of d] {$ <_{\tau} $};
\node (cp) [right=-4pt of dlt, color=Purple] {(c')};

\draw [->,>=stealth,color=PineGreen,thin] ($ (a.south)+(0,-3pt) $) to[out=155,in=25] node[below=-2pt,pos=.9] { $\lrf$ } ($ (b.south)+(0,-3pt) $);
\draw [->,>=stealth,color=PineGreen,thin] ($ (c.south)+(0,-3pt) $) to[out=155,in=25] node[below=-2pt,pos=.9] { $\lrf$ } ($ (d.south)+(0,-3pt) $);

\draw [->,>=stealth,color=CarnationPink,thin] ($ (b.north)+(0,3pt) $) to[out=-135,in=-45] node[above=-2pt,pos=.1] { $\lpo$ } ($ (c.north)+(0,3pt) $);
\draw [->,>=stealth,color=CarnationPink,thin] ($ (c.north)+(0,3pt) $) to[out=-135,in=-45] node[above=-2pt,pos=.1] { $\lpo$ } ($ (d.north)+(0,3pt) $);

\end{tikzpicture}}
	\end{tabular}
	(\hl{W-RWR})
\end{figure}
\setlength{\textfloatsep}{0pt}
Consider the example (\hlref{W-RWR}). Events labeled (a') and (c') are the 
shadow-writes corresponding to the writes labeled (a) and (c), respectively. The
shadow-threads $sth_x(T_1)$  and $sth_x(T_2)$ execute the shadow-write events.
\hlref{W-RWR(i)} shows an execution sequence $\tau$ where
updates to the memory by shadow-writes are illustrated.
\newline
%
As our second central contribution to realize the \mca restriction of
\cmodel model, we define the relation $\emrf$ based on shadow-writes.
%
Let {\small $\lw{\tau}{o}$} represent 
the write corresponding to the 
latest shadow-write of $o$ in $\tau$.
%
\begin{definition}{$\emrf$ relation}{
		For ${e_r \in \reads_\tau}$, its $\emrf$ relation is now defined as: \ext{\newline}
		(general case) $\rf{\tau}{\small \lw{\tau'}{o}}{e_r}$; unless
		\newline
		(special case) $\exists {e_w \in \writes_\tau}$, which is the latest write from 
		$thr(e_r)$ of object $obj(e_r)$ \st
		$\ob{\tau}{\ob{\tau}{\small shw(\lw{\tau'}{o})}{e_w}}{e_r}$
		then 
		$\rf{\tau}{e_w}{e_r}$.
	}
\end{definition}

Intuitively, a read event takes the data from the last write whose shadow-write {\em updated}
the shared memory, unless there is a later write from the same thread.

Consider the execution sequence $\tau$ in example \hlref{W-RWR(ii)}. 
The event (b) reads from (a) (since ${\small \lw{\mbox{a.a'}}{x}}$ = (a)); however, for event
(d) $\lw{\mbox{a.a'.b.c}}{x}$ = (a) but write (c) from same thread occurs 
after (a), thus,  (d) reads from (c).
%
%
%

\noindent {\bf HB relation.} Based on 
$\emrf$ and \cmodel's \textit{Release Sequence} \cite{C11-standard} we tweak
$\emchb$, and its constituent relations \cite{C11-standard} to define
a happens-before relation ($\emhb$) for \ourtechnique.
Note that $\hb{\tau}{}{} \subseteq \ob{\tau}{}{}$.
%
%
For the purpose of recognizing \cmodel behaviors relevant to \mca,
\ourtechnique defines the following relations:
program-order ($\empo$)\footnotemark, 
dependency-ordered-before ($\emdob$), 
synchronizes-with ($\emsw$) and 
inter-thread-happens-before ($\emithb$).
\footnotetext{
	Some events of a thread are not ordered by $\emcsb$ (eg operands
	of ==). We assume a total order ($\empo$) on the events 
	of a thread,
	similar to \cite{norris2013cdschecker,kokologiannakis2017effective}.
}
%

	The relations are formally defined in the table of \reffig{fig:our hb}
	and diagrammatically represented by figures (i) and (ii) on the right of 
	the table
	(note that, the parallel lines ($\parallel$) separate events of 
	different threads).
	The relation $\empo$ is an intra-thread relation that orders the events 
	of a thread in a total-order as shown in \reffig{fig:our hb}(i) and (ii).
	The relation $\emsw$ and $\emdob$ from synchronization between two threads as
	described further. 
	Two strict ordered events when related by the $\emrf$ relation are
	also ordered by $\emsw$, in \reffig{fig:our hb}(ii) 
	$\rf{\tau}{e^{\rel}_{wx}}{e^{\acq}_{rx}}$ where $e^{\rel}_{wx} \in 
	\ordwrites{\moge\rel}_\tau$, $e^{\acq}_{rx} \in \ordreads{\moge\acq}_\tau$ which 
	implies that the two events form a synchronization between their 
	corresponding threads by  forming $\sw{\tau}{e^{\rel}_{wx}}{e^{\acq}_{rx}}$;
	The relation $\emdob$ forms a similar synchronization indirectly \ie instead
	of using an $\emrf$ relation from a strict write event to a strict read event,
	it uses an $\emrf$ relation between a write in the {\em release sequence} 
	(see Appendix~\ref{appdx: formal}) of a strict write and a strict read,
	for example in \reffig{fig:our hb}(i) $\rf{\tau}{e_{wx2}}{e^{\acq}_{rx}}$
	is formed between $e_{wx2} \in \events_{\rs{\tau}{e^{\rel}_{wx1}}}$ and 
	strict read $e^{\acq}_{rx} \in \ordreads{\moge\acq}_\tau$ which forms a
	synchronization between the strict write release head $e^{\rel}_{wx1} \in
	\ordwrites{\moge\rel}_\tau$ and the strict read $e^{\acq}_{rx}$ as
	$\dob{\tau}{e^{\rel}_{wx1}}{e^{\acq}_{rx}}$.

	The synchronizations formed by $\emsw$ and $\emdob$ extend to form the
	inter-thread-hb ($\emithb$) relation by taking a transitive closure of
	$\emsw$, $\emdob$ and $\empo$. As shown in \reffig{fig:our hb}(i)
	$\dob{\tau}{e^{\rel}_{wx1}}{e^{\acq}_{rx}}$ $\implies$
	$\ithb{\tau}{e^{\rel}_{wx1}}{e^{\acq}_{rx}}$.
	Similarly as shown in \reffig{fig:our hb}(ii) 
	$\sw{\tau}{e^{\rel}_{wx}}{e^{\acq}_{rx}}$ $\implies$
	$\ithb{\tau}{e^{\rel}_{wx}}{e^{\acq}_{rx}}$ and
	$\po{\tau}{e_{wy}}{\sw{\tau}{e^{\rel}_{wx}}{\po{\tau}{e^{\acq}_{rx}}}{e_{ry}}}$
	$\implies$ $\ithb{\tau}{e_{wy}}{e_{ry}}$. 
	Further, $\emithb$ is also formed between
	($e_{wy}$, $e^{\acq}_{rx}$) and ($e^{\rel}_{wx}$, $e_{ry}$), the corresponding
	arrows have been skipped in the figure to maintain 	readability.
	Finally, the $\emhb$ relation is formed by taking a union of 
	intra- ($\empo$) and inter- ($\emithb$) thread-hb.

The relation {\em non-racing}-hb ($\emmhb$) relates hb-ordered events
that do not race to access an object either due to $\empo$ ordering
or due to synchronization between their respective threads through
events related by $\emsw$ or $\emdob$.
The relations $\emsw$ and $\emdob$ are also formed between $\fences$,
similar to \cmodel's relations. For brevity we skip the details in the 
paper, please refer Appendix~\ref{appdx: formal}
	for additional details.

%
%


\begin{figure*}[t]
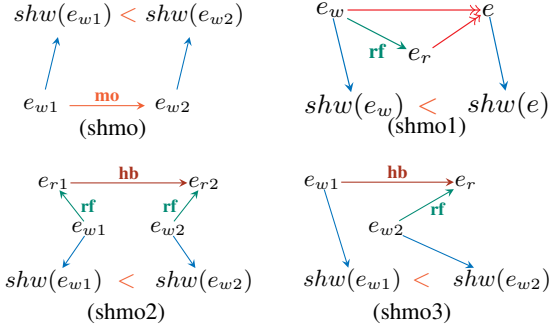

	\small
	\begin{tabular}{|p{0.05\textwidth} p{0.90\textwidth}|}
		\hline
		(\hl{shco}) & $\forall e_r \in \reads_\tau$, if $\tau' = \pre{\tau}{e_r}$
		then, $\exists e_w \in \writes_{\tau'} \st (thr(e_w) = thr(e_r)\ \v\ 
		shw(e_w) \in \modifiers_{\tau'})$,  $e_r$ takes its data from $e_w\ \^\ 
		\nhb{\tau}{e_r}{e_w}$
		\\				
		(\hl{shmo}) & $\forall e',e \in \modifiers_\tau \st obj(e') 
		= obj(e),\ \ob{\tau}{e'}{e} \implies \mo{\tau}{prw(e')}{prw(e)}$
		\\				
		(\hl{shmo1}) & $\forall$ $e_w \in \writes_\tau$, $e_r \in \reads_\tau$, 
		$e$ $\in$ $\events_\tau$, \st $thr(e)$ $\neq$ $thr(e_w)$,
		$\mhb{\tau}{e_w}{e}$ $\v$ $\mhb{\tau}{\rf{\tau}{e_w}{e_r}}{e}$ $\implies$
		$\ob{\tau}{shw(e_w)}{e}$ (if $e \nin \writes_\tau$) $\^$
		$\ob{\tau}{shw(e_w)}{shw(e)}$ (if $e \in \writes_\tau$)
		\\		
		(\hl{shmo2}) & $\forall e_{r1}, e_{r2} \in \reads_\tau \st 
		\hb{\tau}{e_{r1}}{e_{r2}}$ if $\exists e_{w1},e_{w2} \in \writes_\tau
		\st \rf{\tau}{e_{w1}}{e_{r1}} \^ \rf{\tau}{e_{w2}}{e_{r2}}$ where 
		$e_{w1} \neq e_{w2}$ then $\ob{\tau}{shw(e_{w1})}{shw(e_{w2}})$
		\\				
		(\hl{shmo3}) & $\forall e_{w1},e_{w2} \in \writes_\tau, e_r \in 
		\reads_\tau \st \hb{\tau}{e_{w1}}{e_r} \^ \rf{\tau}{e_{w2}}{e_r}$ 
		where $e_{w1} \neq e_{w2}$, $\ob{\tau}{shw(e_{w1})}{shw(e_{w2})}$
		\\				
		(\hl{shrmo}) & $\forall e \in \events_\tau, act(e) =$ rmw,
		$\exists \rf{\tau}{e_w}{e} \st \mo{\tau}{e_w}{e}$ $\^$
		$\nexists e_w' \st \mo{\tau}{\mo{\tau}{e_w}{e_w'}}{e}$
		\\
		\hline
	\end{tabular}
	\caption{Coherence rules for traces in \ourtechnique}
	\label{fig: shmo rules}
	\vspace{-1em}
\end{figure*}	

\noindent {\bf Coherence rules.}
Read value coherence for \ourtechnique is a derivative of $\emrf$ and is
interpreted as:
\begin{itemize}
	\item (\hlref{shco}) a read event must take its value either from (i) a valid write event
	whose shadow-write does not occur after the read, or (ii) from a valid 
	write event of the same thread that does not occur after it	
	(refer \reffig{fig: shmo rules} for formal definition).
\end{itemize}

\begin{figure}[t]
	\centering
	\begin{tabular}{cc}
		\resizebox{.4\columnwidth}{!}{\tikzset{every picture/.style={line width=0.75pt}} 
\begin{tikzpicture}[x=1em,y=1em,yscale=1,xscale=1]
	
\tikzstyle{every node}=[font=\small]
\node (shew1) {$ shw(e_{w1})$};
\node (less) [right=-5pt of shew1,color=RedOrange] {$<$};
\node (shew2) [right=5pt of shew1] {$ shw(e_{w2}) $};
\node (ew1) [below left=20pt and -25pt of shew1] {$ e_{w1} $};
\node (ew2) [below left=20pt and -25pt of shew2] {$ e_{w2} $};
\node (shmo) [below right=-5pt and 0pt of ew1] {(shmo)};

\draw [->,>=stealth,thin,color=RoyalBlue] (ew1) -- (shew1);
\draw [->,>=stealth,thin,color=RoyalBlue] (ew2) -- (shew2);
\draw [->,>=stealth,thin,color=RedOrange] (ew1) -- node[midway,above=-2pt,color=black,font=\scriptsize] { $\lmo$ } (ew2);

\end{tikzpicture}} &
		\resizebox{.4\columnwidth}{!}{\tikzset{every picture/.style={line width=0.75pt}} 
\begin{tikzpicture}[x=1em,y=1em,yscale=1,xscale=1]
	
\tikzstyle{every node}=[font=\small]
\node (ew) [inner sep=0] {$e_w$};
\node (e) [right=45pt of ew,inner sep=0] {$ e $};
\node (er) [below right=10pt and 20pt of ew,inner sep=0] {$ e_r $};
\node (shew) [below right=25pt and -15pt of ew,inner sep=0] {$ shw(e_w) $};
\node (less) [right=0pt of shew,color=RedOrange] {$<$};
\node (she) [below right=25pt and -10pt of e,inner sep=0] {$ shw(e)$};
\node (shmo1) [below right=-5pt and -10pt of shew,font=\footnotesize] {(shmo1)};

\draw [->,>=stealth,thin,color=RoyalBlue] (ew) -- (shew);
\draw [->,>=stealth,thin,color=RoyalBlue] (e) -- (she);
\draw [->>,thin,color=Red] (er) -- (e);
\draw [->>,thin,color=Red] (ew) -- (e);
\draw [->,>=stealth,color=PineGreen,thin] (ew) -- node[below,midway,color=black,font=\scriptsize] { $\lrf$ } (er);

\end{tikzpicture}} \\
		\resizebox{.4\columnwidth}{!}{\tikzset{every picture/.style={line width=0.75pt}} 
\begin{tikzpicture}[x=1em,y=1em,yscale=1,xscale=1]
\tikzstyle{every node}=[font=\small]
\node (er1) [inner sep=0] {$e_{r1}$};
\node (er2) [right=45pt of er1,inner sep=0] {$ e_{r2} $};
\node (ew1) [below right=12pt and 0pt of er1,inner sep=0] {$ e_{w1} $};
\node (ew2) [below left=12pt and 0pt of er2,inner sep=0] {$ e_{w2} $};
\node (shew1) [below =30pt of er1,inner sep=0] {$ shw(e_{w1}) $};
\node (shew2) [below =30pt of er2,inner sep=0] {$ shw(e_{w2}) $};
\node (less) [right=1.5pt of shew1,color=RedOrange] {$<$};
\node (shmo2) [below right=0pt and -10pt of shew1] {(shmo2)};

\draw [->,>=stealth,thin,color=RoyalBlue] (ew1) -- (shew1);
\draw [->,>=stealth,thin,color=RoyalBlue] (ew2) -- (shew2);

\draw [->,>=stealth,color=PineGreen,thin] (ew2) -- node[left,midway,color=black,font=\scriptsize] { $\lrf$ } (er2);
\draw [->,>=stealth,color=PineGreen,thin] (ew1) -- node[right,midway,color=black,font=\scriptsize] { $\lrf$ } (er1);
\draw [->,>=stealth,color=Mahogany,thin] (er1) -- node[above=-2pt,midway,color=black,font=\scriptsize] { $\lhb$ } (er2);

\end{tikzpicture}} &
		\resizebox{.4\columnwidth}{!}{\tikzset{every picture/.style={line width=0.75pt}} 
\begin{tikzpicture}[x=1em,y=1em,yscale=1,xscale=1]
\tikzstyle{every node}=[font=\small]
\node (ew1) [inner sep=0] {$e_{w1}$};
\node (er) [right=45pt of ew,inner sep=0] {$ e_r $};
\node (ew2) [below right=12pt and 10pt of ew1,inner sep=0] {$ e_{w2} $};
\node (shew1) [below right=30pt and -15pt of ew1,inner sep=0] {$ shw(e_{w1}) $};
\node (less) [right=0pt of shew1,color=RedOrange] {$<$};
\node (shew2) [below right=30pt and -10pt of er,inner sep=0] {$ shw(e_{w2})$};
\node (shmo3) [below right=0pt and -15pt of shew1] {(shmo3)};

\draw [->,>=stealth,thin,color=RoyalBlue] (ew1) -- (shew1);
\draw [->,>=stealth,thin,color=RoyalBlue] (ew2) -- (shew2);
\draw [->,>=stealth,color=Mahogany,thin] (ew1) -- node[above=-2pt,midway,color=black,font=\scriptsize] { $\lhb$ } (er);
\draw [->,>=stealth,color=PineGreen,thin] (ew2) -- node[below=-2pt,pos=.7,color=black,font=\scriptsize] { $\lrf$ } (er);

\end{tikzpicture}}
	\end{tabular}
\end{figure}

The (\hlref{shco}) rule disallows the discussed behavior
in \hlref{IRIW} (\refsection{sec:intro}).
%
We introduce a set of \ourtechnique-{\em mo rules} based on which 
\ourtechnique 
determines the order of occurrence of shadow-writes of an object and
help determine $\emmo$; they are formally defined in \reffig{fig: shmo
	rules} and represented diagrammatically in \reffig{fig: shmo figs}.
\begin{itemize}
	\item (\hlref{shmo}): writes $e_{w1}$, $e_{w2}$ are mo-ordered if
	$shw(e_{w1})$ occurs before $shw(e_{w2})$.
	\item (\hlref{shmo1}): if a write $e_w$ is $\emmhb$ or $\emrf$ ordered 
	with event $e$ from another
	thread, then either $shw(e_w)$ 
	must occur before $shw(e)$ (if $e \in \writes$) or before $e$ (if 
	$e \nin \writes$);
	\item (\hlref{shmo2}): if a read $e_{r1}$ is hb-ordered before another read $e_{r2}$, then
	the shadow-write of ${e_{r1}}$'s source must occur before the  
	shadow-write of ${e_{r2}}$'s source; 
	\item (\hlref{shmo3}): shadow-write of a read's source must occur after
	shadow-writes of all writes hb-ordered before the read.
	\item (\hlref{shrmo}): To ensures atomicity, 
	each rmw event must read-from the {\em immediately}
	ordered before event in the modification order. 
\end{itemize}
The above rules assist \ourtechnique in constructing coherent \cmodel sequences.
%
Notably \ourtechnique also maintains a total order relation $\emto$ 
on \sc events. It does so
in the following way:
(i) all \sc events in $\empo$ are also in $\emto$ relation, and
(ii) \sc ordered events from different 
threads are in $\emto$ by their occurrence order, except write events
that are in $\emto$ by the occurrence order of their shadow-write.
Based on $\emto$ coherence on \sc events is maintained by rule
(\hlref{shto}): all \sc ordered events must form a total-order ($\emto$)
\wrt $\emmo$ and $\emhb$.
%
%
%
\newline
A maximal sequence, $\tau$, and the associated
$\emhb$
are coherent and represent a \ourtechnique
trace if
(\hlref{shmo}),  (\hlref{shmo1}), (\hlref{shmo2}), (\hlref{shmo3}),
(\hlref{shrmo}), (\hlref{shco}) and (\hlref{shto}) are satisfied by $\tau$.
Finally, through Theorem~\ref{thm:coherence}, we demonstrate that 
traces generated by \ourtechnique are indeed coherent under \cmodel. 
%

\vspace{-1em}
\theorem{ $\forall \tau,\ \hb{\tau}{}{}\
	\subseteq\ \chb{\tau}{}{}$
}{coherence}
\noindent
%
\proof{
	The coherence rules of \ourtechnique satisfy the coherence rules
	of \cmodel \cite{C11-standard}. 
	Thus, \ourtechnique traces are coherent \cmodel traces.
	See Appendix~\ref{proof: coherence} for formal proof.
}

\section {\cmodel-\mca aware \sdpor} \label{sec:algo}
\ourtechnique explores all relevant program
behaviors for detecting safety assertion violations as well as 
non-atomic (\na) data races.
Central to \ourtechnique is  \sdpor (Algorithm 1 of \cite{abdulla2014optimal}),
which is a near-optimal improvement over 
DPOR \cite{flanagan2005dynamic}. 
%
%
It is noteworthy that \sdpor algorithm  used in \ourtechnique 
is {\em as is}, \ie, 
without any modification. This was feasible because of several reasons:
\begin{itemize}
	\item our design of a valid happens-before relation (Theorem~\ref{thm:coherence}) 
	for restricting \cmodel  under \mca 
	is directly pluggable in \sdpor, 
	\item our proposal of shadow-threads, 
	shadow-writes makes it possible to avoid reordering instructions
	from a thread during exploration and rely on interleaving model of
	computation alone, and
	\item the parallel composition rule \hlref{(parcom)}(\refsection{sec:mca}) 
	satisfies the requirement
	of \sdpor that only one thread executes at a time.
\end{itemize}
\Sdpor is a non-chronological depth-first search of a directed acyclic
graph of execution states. Much like the quintessential 
DPOR \cite{flanagan2005dynamic}, \sdpor maintains set of events that 
should be explored at each state and a set of {\em sleeping} threads. 
However, unlike the classical DPOR, \sdpor computes 
much compact set of possible starts from a state than {\em persistent sets} 
\cite{mc-book}. We invite the reader to refer to \cite{abdulla2014optimal} 
for \sdpor details.  
\noindent {\bf Shadow-threads and shadow-writes:}
%
The shadow-threads introduced in our technique are handled in the following way
so that \sdpor algorithm can be used {\em as is}:
whenever an event $e \in \writes$ of thread $\tid{i}$ is executed from a 
state $\s{\tau}$, a corresponding shadow-write event $shw(e)$ is generated
and added to the shadow-thread $\tid{si}$ corresponding to $obj(e)$.
Similar to program threads, execution of an enabled shadow-write 
$shw(e)$ of a shadow-thread $\tid{si}$ from a state $\s{\tau}$ enables the 
next event of $\tid{si}$ at state $\s{\tau.shw(e)}$.

%
The shadow-writes, as remarked before, enable reordering through 
interleaving. Note, however, that a
shadow-write is created only after a corresponding program write
has occurred. As a result shadow-writes simulate the reordering
of program writes with {\em later} events from the same
thread. An important  question that arises is: how \ourtechnique covers
the case of a {\bf program write reordering with an {\em earlier}
	program event} from the same thread?
%
\ourtechnique implicitly assumes that the writes are at the earliest 
location in the program possible (where earlier refers to a lower event 
index). To meet this requirement \ourtechnique performs a static  
{\em \pretransformation} transformation from input  program \program to 
\programhat.

\noindent
{\tt \hl{\Pretransformation}:} 
The transformation rules for each thread sequence $\tid{i}{:}\tau$ of the
original program, \program, are: 

If there exists a corresponding thread sequence
$\tid{i}{:}\tau'$ of the transformed program then,
\begin{itemize}[\IEEEsetlabelwidth{ewt1}]
	\item [ewt1] 
	$\events_{\tid{i}{:}\tau}$ = $\events_{\tid{i}{:}\tau'}$
	(\ie, same event sets but their order of occurrence may vary);
	
	\item [ewt2] if $\exists e_1, e_2$ \st
	$\ob{\tid{i}{:}\tau}{e_1}{e_2}$ $\wedge$
	$\ob{\tid{i}{:}\tau'}{e_2}{e_1}$, then $e_2$ $\in$ $\writes$
	$\^$ $(\nexists e_3 \in \events_{\tid{i}{:}\tau}$ \st
	$\ob{\tid{i}{:}\tau}{e_1 \le_{\tid{i}{:}\tau} e_3}{e_2}$ $\^$
	$(\dep{e_3}{e_2}$ $\v$ $\noup{e_3, e_2}))$ (\ie, $e_2$ is a
	write and can reorder above $e_1$ only if there is no intervening
	$e_3$ that either introduces program dependency with $e_2$ or creates
	upward reordering restriction).
\end{itemize}

We show through
Theorem~\ref{thm:Phat semantic preserving} that \pretransformation
transformation does not alter the semantics of \program. 
As a result, instead of \program, the transformed program \programhat is provided
to the \sdpor algorithm as input.
%

\setlength{\textfloatsep}{0pt}
\begin{table}[t]
	\caption{Comparative Results on litmus tests}
	\label{tab: compare tests}
	\footnotesize
	\setlength{\tabcolsep}{3pt}
	\resizebox{\columnwidth}{!}{%
		\begin{tabular}{|l|r r r|r r r|r r r|r r r|}
			\hline
			\multirow{2}{*}{Test} & \multicolumn{3}{c|}{\ourtechnique} & 
			\multicolumn{3}{c|}{\cds} & \multicolumn{3}{c|}{\genmc} &
			\multicolumn{3}{c|}{\hmc} \\\cline{2-13}
			& M & N & Time & M & N & Time & M & N & Time &
			M & N & Time \\
			\hline
			
			WRC+addrs(7) & 7 & 0 & 0.03s & 7 & 1 & 0.01s & 7 & 1 & 0.02s & 7 & 1 & 0.03s \\  
			WR-ctrl(4) & 7 & 0 & 0.03s & 4 & 2 & 0.01s & 4 & 2 & 0.02s & 4 & 2 & 0.02s \\
			Z6+poxxs(4) & 18 & 0 & 0.12s & 14 & 4 & 0.01s & 4 & 4 & 0.03s & 4 & 4 & 0.03s \\ 
			IRIW+addrs(15) & 15 & 0 & 0.07s & 15 & 1 & 0.01s & 15 & 1 & 0.02s & 15 & 1 & 0.02s \\
			WW+RR(15) & 96 & 0 & 0.53s & 15 & 66 & 0.02s & 15 & 66 & 0.02s & 15 & 66 & 0.02s \\
			\hline	
			\multicolumn{13}{r}{M: $\#$\mca sequences, N: $\#$non-\mca sequences}\\
		\end{tabular}
	}
\end{table}

\setlength{\textfloatsep}{0pt}
\begin{table}[t]
	\caption{Comparative Results on benchmarks}
	\label{tab: compare benchmarks}
	\footnotesize
	\setlength{\tabcolsep}{3pt}
	\resizebox{\columnwidth}{!}{%
		\begin{tabular}{|l|r r|r r|r r|r r|}
			\hline
			\multirow{2}{*}{Test} & \multicolumn{2}{c|}{\ourtechnique} & 
			\multicolumn{2}{c|}{\cds} & \multicolumn{2}{c|}{\genmc} &
			\multicolumn{2}{c|}{\hmc} \\\cline{2-9}
			& $\#$Seq & Time & $\#$Seq & Time & $\#$Seq & Time &
			$\#$Seq & Time \\
			\hline
			mutex & 5 & 0.02s & 2-NVs & 0.01s & NV & 0.03s & NV & 0.02s \\
			peterson & 13 & 0.15s & 666-NVs & 2.73s & NV & 0.02s & NV & 0.02s \\
			RW-lock & 246 & 0.52s & 193-NVs & 0.38s & NV & 0.02s & NV & 0.04s \\
			spinlock & 506 & 16.98s & TO & - & NV & 0.08s & NV & 0.15s \\\hline
			fibonacci-2 & 667 & 5.57s & TO & - & NV & 0.04s & NV & 0.03s \\
			fibonacci-3 & 10628 & 2m14s & TO & - & NV & 0.06s & NV & 0.07s \\
			fibonacci-4 & 92421 & 56m21s & TO & - & NV & 0.13s & NV & 0.31s \\\hline
			counter-5 & 3599 & 39.78s & 25-NVs & 0.31s & NV & 0.06s & NV & 0.03s \\
			counter-10 & 55927 & 12m53s & 100-NVs & 9.21s & NV & 0.05s & NV & 0.07s \\
			counter-15 & TO & - & 225-NVs & 50.31s & NV & 0.11s & NV & 0.16s \\\hline
			flipper-5 & 2489 & 20.19s & 201-NVs & 3.26s & NV & 0.03s & NV & 0.04s \\
			flipper-10 & 96737 & 6m12s & TO & - & NV & 0.04s & NV & 0.02s \\
			flipper-15 & TO & - & TO & - & NV & 0.03s & NV & 0.03s \\\hline
			prod-cons-10 & 9373 & 1m23s & TO & - & NV & 0.04s & NV & 0.04s \\
			prod-cons-15 & 38593 & 6m46s & TO & - & NV & 0.02s & NV & 0.02s \\
			prod-cons-20 & 109838 & 20m28s & TO & - & NV & 0.02s & NV & 0.02s \\
			\hline	
			\multicolumn{9}{l}{x-NVs: `x' number of non-\mca violations (if $\#$violations reported)}\\
			\multicolumn{9}{l}{NV: non-\mca violation (if analysis halts at first violation, $\#$NVs not known)}\\
			\multicolumn{9}{l}{TO: timeout (60m)}\\
			\multicolumn{9}{l}{{\em Additionally, \ourtechnique detected \na races in tests} mutex {\em and} counter.}
		\end{tabular}
	}
\end{table}

\theorem{\Pretransformation (\program to \programhat) is semantics preserving.}
{Phat semantic preserving}
\proof{
	We show that rules ewt1 and ewt2 are compliant with 
	semantic preserving reordering rules spr1-3 
	defined for \mca model in \refsection{sec:mca}); 
	Refer the Appendix~\ref{proof: soundness} for details.
}

Via Theorem~\ref{thm:coherence}, we established that any trace of \program 
that \ourtechnique examines is a valid \cmodel trace. However, to show
that \ourtechnique explores precisely the \mca traces of \cmodel program, 
we present Theorem~\ref{thm:traces eq}.

\vspace{-1em}
\theorem{\ourtechnique traces are equivalent to \cmodel traces valid over \mca}
{traces eq}
\proof{
	{\em Case $\leftarrow$}:
	$\forall e',e \st \neg(\eventreorder{e'}{e}) \implies 
	dep(e',e)$ $\v$ $\noup{e',e}$ $\implies$ if $\ob{\tid{i}{:}\tau}{e'}{e}$ then
	$\ob{\tid{i}{:}\tau'}{e'}{e}$. 
	Further, if $e' {\in} \writes_\tau$, $e$ can observe its effect.
	(by construction of $\emrf$ and shadow-threads).
	Thus, reordering restricted by \cmodel over \mca is also 
	restricted by \ourtechnique. {\em inf} (i).
	As \pretransformation is semantic preserving, thus reordering allowed
	under \ourtechnique is allowed under \mca. {\em inf} (ii).
	$e_w \in \writes$ along with $shw(e_w)$ perform 
	\hlref{(w-issue)} and \hlref{(w-update)} while preserving semantics
	(using (\hlref{shco})).
	$e_r \in \reads$ perform \hlref{(r-shared)}.
	\Sdpor algorithm ensures \hlref{(parcom)}. {\em inf} (iii).
	Thus, from {\em inf} (i), (ii), (iii),
	traces of \ourtechnique have equivalent \cmodel \mca 
	traces. 
	
	{\em Case $\rightarrow$}:
	If $\ob{\tid{i}{:}\tau}{e'}{e}$ and $\ob{\tid{i}{:}\tau'}{e'}{e}$ then 
	$\dep{e'}{e} \v \noup{e',e}$
	$ \implies \neg(\eventreorder{e'}{e})$ and 
	reordering of $e'$ and $e$ is not supported by \cmodel.
	Hence, reordering restricted by \ourtechnique is also restricted 
	by \cmodel under \mca. {\em inf} (iv).
	$\forall \eventreorder{e'}{e}$ $\implies$ 
	$obj(e') \neq obj(e)$ and thus effect of $e',e$ can interleave
	under \ourtechnique.
	%
	Hence, reordering allowed under \mca are allowed by 
	\ourtechnique. {\em inf} (v).
	\hlref{(r-shared)} is performed by $\reads$, \hlref{(w-issue)} and 
	\hlref{(w-update)} by $\writes$ and $\modifiers$,
	\hlref{(parcom)} is ensured by \sdpor algorithm.
	{\em inf} (vi).
	Thus, from {\em inf} (iv), (v), (vi) \cmodel \mca traces are valid 
	\ourtechnique traces.
}

\noindent
{\bf Design complexity:}
	The complexity of \sdpor is $O(T^2|\events|^2S)$, 
	where $T$ is the number of threads and $S$ 
	is the number of  sequences explored.
	The relation $\emhb$ has the same computational complexity as 
	in the original \sdpor work, \ie $O(|\events|^2)$. 
	Addition of shadow-threads, however, 
	makes the worst-case complexity of \ourtechnique 
	$O(|\objects|^2T^2|\events|^2S)$.

%
\section{Experimental Validation} \label{sec:results}

\noindent {\bf Implementation details.} 
We present a prototype implementation to experimentally validate 
\ourtechnique technique. 
%
The implementation is built on \rinspect \cite{zhang2015dynamic}.
\ourtool takes a \cmodel program as input.
The input program \program
is statically transformed to \programhat by the \pretransformed transformation
(\refsection{sec:algo}). 
Further, \programhat 
is instrumented using LLVM to recognize newly enabled events dynamically
during execution of a sequence.
The events are communicated to 
\ourtechnique  scheduler, which orchestrates the 
order of execution of events using \sdpor algorithm. 
%
%
\ourtool re-runs \programhat for every maximal sequence explored.
After analyzing the input program \program, \ourtechnique reports
assert violations if any.
According to \cmodel standard  if the order of occurrence 
of a pair of \na ordered events
can potentially be reversed in a trace, then the 
behavior of the trace is {\em  undefined}.
Such a behavior can defy the
coherence specification (\hlref{shco}), (\hlref{co}) and produce invalid values. 
Thus, \ourtechnique
also reports {\em data races on non-atomic memory accesses}.

\setlength{\textfloatsep}{0pt}
\setlength{\tabcolsep}{3pt}
\begin{table}[t]
	\begin{minipage}{0.4\columnwidth}
		\centering
		\caption{\mca tests}
		\label{tab: mca tests}
		\scriptsize
		\begin{tabular}{|l|r r|}
			\hline
			Test & $\#$Seq & Time \\
			\hline
			
			CoRR(3) & 3& 0.02s \\
			CO-RSDWI(3) & 6 & 0.02s \\
			R+fn+fn(2) & 5 & 0.02s \\
			RSDWI(6) & 22 & 0.11s \\
			WRR+2W(12) & 29 & 0.12s \\
			Luc17(12) & 12 & 0.07s \\
			Luc10(3) & MV & 0.02s \\
			S-popl(3) & MV & 0.01s \\
			\hline	
			\multicolumn{3}{r}{MV: \mca violation(s) detected}\\
			\multicolumn{3}{r}{}\\
			\multicolumn{3}{r}{}
		\end{tabular}
	\end{minipage}
	\begin{minipage}{0.6\columnwidth}
		\centering
		\caption{\cmodel tests}
		\label{tab: c11 tests}
		\scriptsize
		\begin{tabular}{|l|r r|r r|}
			\hline
			Test & $\#$Seq & Time & race? & $\#$Rseq \\
			\hline
			
			simple-sw(3) & 3 & 0.006s & Y & 2 \\
			simple-ithb(4) & 4	& 0.034s & Y & 2 \\
			RS-blk(10) & MV & 0.07s & Y & 8 \\
			CSE-no-blk(8) & 12 & 0.158s & N & - \\
			no-fence-sync(5) & MV & 0.054s & Y & 5 \\	
			fib-no-assert & 26 & 0.14 & N & - \\
			fmax-cas & 31 & 0.21s & N & - \\
			flipper & 1628 & 9.29s & N & - \\
			\hline	
			\multicolumn{5}{l}{MV: \mca violation(s) detected} \\
			\multicolumn{5}{l}{race?: Does the test have a race on \na events?} \\
			\multicolumn{5}{l}{$\#$Rseq: number of \na races detected by \ourtechnique}
		\end{tabular}
	\end{minipage}
\end{table}

\noindent {\bf Experiment details.} 
We performed tests to validate  correctness \wrt
\mca using diy7 family of litmus tests 
\cite{diy7} (sample listed in \reftab{tab: mca tests}). 
To test \cmodel coherence, we synthesized 56  litmus
tests relevant to the \cmodel coherence rules (eg. row 1-4, 
\reftab{tab: c11 tests}) and borrowed multi-threaded benchmarks 
from the SV-Comp benchmark suite \cite{svcomp} (eg. row 5-8, \reftab{tab: c11 tests}). 
We 
remodeled them for \cmodel with the use of atomic data types and 
\cmodel memory orders.
%
We recorded time (column `Time')
and the number of maximal sequences explored 
(column `$\#$Seq'), which includes {\em at least}
one execution corresponding to each trace and
(possibly) few redundant executions owing to the non-optimal 
nature of the underlying \sdpor algorithm.
Further, if a test contains \na races (column `race?') then we 
report the number of maximal sequences that contain \na race(s)
(column `$\#$Rseq').
%
%

To demonstrate the effectiveness of \ourtechnique, we used 
litmus tests and benchmarks from the SV-Comp suite that produce
a {\em strict subset} of \cmodel behaviors when restricted to \mca.
We compared the outcome of such tests on \ourtool with \sota 
stateless model checking tools for \cmodel (and its variants) namely \cds 
\cite{norris2013cdschecker} and \genmc 
\cite{kokologiannakis2019model}; and hardware model checker
(\hmc) \cite{kokologiannakis2020hmc}. 

\noindent {\bf Results' analysis.} 
\reftab{tab: compare tests} shows the results of comparative 
study on litmus tests that would show additional behaviors on
non-\mca model. The table contains small tests with 15 or less
traces.
The number of valid \cmodel traces
under \mca have been shown 
in bracket accompanying the name of the test. 
For instance, `WRC+addrs(7)', shows that
the test `WRC+addrs' has 7 valid \mca 
\cmodel traces . These numbers have been manually computed. 
We have
reported the number of \mca sequences (column `M')   
and the number of non-\mca sequences (column `N') for each of
the techniques \ourtechnique, \cds, \genmc and \hmc, 
along with the 
time taken by the techniques for performing their analysis.

For larger benchmarks we have used assert
statements to catch non-\mca sequences and used `NV' to indicate
assert violation(s) in non-\mca sequences. The results are shown in 
\reftab{tab: compare benchmarks}. An `NV' result implies that 
the benchmark may have a legitimate violaton under {\cmodel model but
	not under \mca}.

\cds reports all sequences explored
including ones with assert violation, thus, for \cds the
collected number of non-\mca violations  have been reported 
as `x-NVs', indicating `x' number of assert violations in
non-\mca sequences. 
\genmc and \hmc halt at the first detection of violation, thus,
no such information is available for reporting. 
Hence, for \genmc and \hmc we have simply written `NV'. 
As a consequence, the time reported for \genmc and \hmc is 
the time to encounter the first assert violation and is 
therefore much lower than the time reported by \cds and 
\ourtechnique.
Due to such difference in tool design, the reported time of 
analysis is incomparable and has been reported only for
reference.
The value `TO' indicates timeout set for 60 minutes.

Finally, we re-emphasize that the techniques \cds and \genmc are
designed for \cmodel (or its variants) and \hmc is for a collection of hardware models
subsuming \mca.  Naturally, these techniques explore a larger set of
traces, and 
the non-\mca violation(s) reported by them are  indeed
true violations under their respective models. 
However, some of the violations reported by them 
may never manifest on the underlying
architecture.
We can observe from \reftab{tab: compare tests} and 
\reftab{tab: compare benchmarks} that benchmarks can 
produce hundreds of assert violations that may not be reproducible
on an actual architecture. 
Thus, a precise technique for \mca such as \ourtechnique can be useful.

%
\section{Related Work} \label{sec:related-work}
\noindent {\bf Stateless model checking:}
Stateless model checking (SMC) with
DPOR \cite{flanagan2005dynamic} has been used for
(SC) \cite{abdulla2014optimal}\cite{albert2017context}\cite{Chalupa:2017}\cite{Rodrguez2015Unfolding} \cite{NguyenRSCP18}
and weak memory models (WMM)
TSO, PSO~\cite{abdulla2015stateless}\cite{zhang2015dynamic},
Power \cite{abdulla2016stateless} and \cmodel \cite{norris2013cdschecker}. 
The techniques \cite{kokologiannakis2017effective}\cite{kokologiannakis2019model}\cite{abdulla2018optimal}
have proposed SMC for variants of \cmodel and~\cite{kokologiannakis2020hmc} 
for a superset of
architectural memory models (including \mca model).

\noindent{\bf Symoblic Analyses:} 
Symbolic and predictive trace analysis is investigated 
in \cite{forejt2014precise}\cite{gupta2015succinct}\cite{WangKGG09} and
has been 
applied to the verification of MPI
programs~\cite{KhannaSRP18}.
%
%
Static analysis  
using thread modular analysis 
or abstract interpretation have been proposed
under SC \cite{flanagan2002thread}\cite{kusano2016flow} and 
WMM \cite{kusano2017thread}\cite{suzanne2018relational}.
While these techniques are sound, they may suffer from false alarms.
Recent works have also investigated bounded model checking
under loop and {\em view} bounds to analyze WMM
\cite{abdulla2019verification}\cite{ponce2020dartagnan}.

\section{Concluding Remarks} \label{sec:conclusion}

We present \ourtechnique, a dynamic verifier to analyze \cmodel
program traces valid under \mca model for assertion violations 
na data races.
%
The technique is shown to be sound and precise.  The empirical results
demonstrate the utility of \ourtechnique over existing techniques
for \cmodel.

\noindent
{\bf Future Work:}
In future we would like to explore the extensions of our work
to reactive systems and
include richer program constructs such as locks and memory barriers.
Another area of possible investigation would be to combine current work with
symbolic trace verification so as to avoid 
re-runs of the input program.

\begin{footnotesize}
\bibliographystyle{IEEEtran}
\bibliography{sections/references}
\end{footnotesize}

\onecolumn
\setcounter{theoremcounter}{-1}
\begin{appendices}
\section{Formal Definitions} \label{appdx: formal}
Given a sequence $\tau$, the sequence $\tau.e$ is obtained by extending $\tau$ by an event $e$. 
For a sequence $\tau = e.\tau'$, $\hd{\tau}$ and $\tl{\tau}$ denote 
$e$ and $\tau'$, respectively.

\begin{definition}{Subword relation}{
  A binary relation  $\subseq{}{}$ is a  subword relation s.t.
  $\subseq{\tau_1}{\tau_2}$ holds for two trace prefixes $\tau_1$ and $\tau_2$ only when
  $\tau_1$ is a subword of $\tau_2$. }
\end{definition}

\noindent {\em Example:} Given $\tau = e_1.e_2.e_3.e_4$,  sequence 
$\subseq{e_1.e_3}{\tau}$ is an example of subword of $\tau$.

\begin{definition}{Release Sequence}{
		 Given $e \in \ordwrites{\moge\rel}_\tau$, 
		$ \rs{\tau}{e}$ is a release sequence starting at $e$
		if $\rs{\tau}{e}$ is the maximal sequence $r$ \st 
		$\subseq{r}{\tau}$
		$\st \forall e'\in \events_r, e' \in \writes_\tau$
		$\^ \hd{r} = e \^ 
		\nexists e''\in \ordwrites{\molt\rel}_{\tl{r}}$ 
		$\st thr(e'') \neq thr(e) \^$
		$act(e'') \neq$ rmw.
}
\end{definition}

	\begin{figure}[!h]
		\centering
		\includegraphics[scale=0.6]{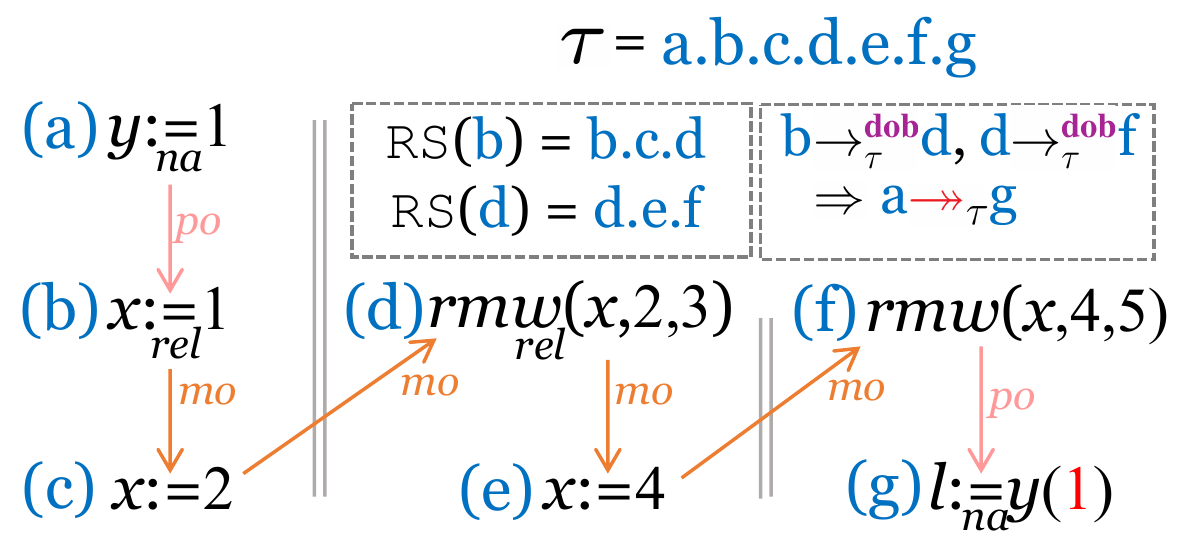}
		\caption{Example Release Sequence}
		\label{fig:rs}
	\end{figure}

\begin{definition}{Downward restriction}{
For any event $e \in \ordwrites{\moge \rel}$ and for any event 
$e' \in \events_{\tau} \wedge thr(e) = thr(e')$, the downward 
restriction $\nodown{e, e'}$ defines a predicate which holds 
when $\ob{\tau}{e'}{e}$. 

}
\end{definition}

\begin{definition}{Upward restriction}{
For any event $e \in \ordreads{\moge \acq}$ and for any event 
$e' \in \events_{\tau} \wedge thr(e) = thr(e')$, the upward 
restriction $\noup{e, e'}$ defines a predicate which holds 
when $\ob{\tau}{e}{e'}$. 

}
\end{definition}

The $\emsw$ relation can also be formed by \cmodel fences as described
below.
\begin{figure}[h]
	\centering
	\begin{tabular}{|p{0.2\textwidth}|p{0.2\textwidth}|p{0.2\textwidth}|}
		\multicolumn{3}{c}{given $f_\rel {\in} \ordfences{\moge \rel}_\tau,$ 
			$f_\acq {\in} \ordfences{\moge \acq}_\tau$,
			$e_{wx} {\in} \ordwrites{\moge \na}_\tau,$ 
			$e_{rx} {\in} \ordreads{\moge \na}_\tau$
			and $\rf{\tau}{e_{wx}}{e_{rx}}$} \\
		\hline
		\includegraphics[scale=0.6]{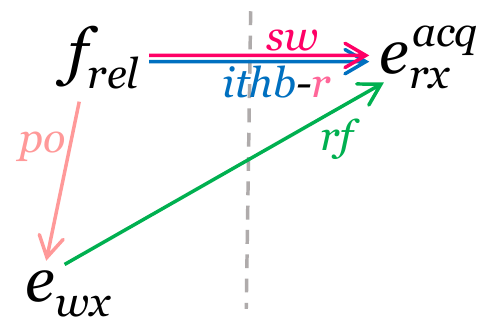} &
		\includegraphics[scale=0.6]{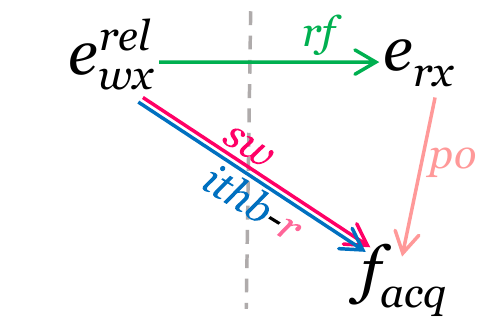} &
		\includegraphics[scale=0.6]{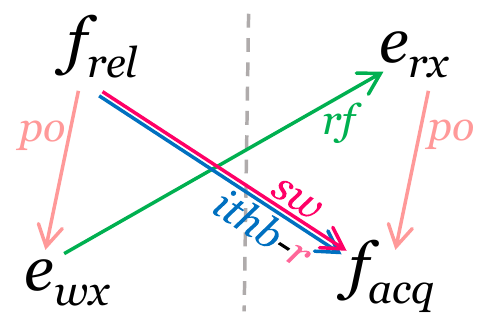} \\
		$\sw{\tau}{f_\rel}{e_{rx}}$ if \newline
		$ord(e_{rx}) \moge \acq,$ $\po{\tau}{f_\rel}{e_{wx}}$  &
		$\sw{\tau}{e_{wx}}{f_{\acq}}$ if\newline
		$ord(e_{wx}) \moge \rel,$ $\po{\tau}{e_{rx}}{f_\acq}$ &
		$\sw{\tau}{f_\rel}{f_\acq}$ if\newline
		$\po{\tau}{f_\rel}{e_{wx}},$ $\po{\tau}{e_{rx}}{f_\acq}$\\\hline
		\multicolumn{3}{c}{(\hl{fence-sw})}
	\end{tabular}
	\label{fig:fences}
\end{figure}

\newpage
\section{Proof of Coherence} \label{proof: coherence}
\noindent
Let $val(\tau,e)$ represent the value read by $e$ if $e \in \reads_\tau$
or the value written by $e$ if $e \in \writes_\tau$ or $e \in \modifiers_\tau$.
\newline

\noindent
Let (ithb-reversible) $\emithbr$ represent the $\emithb$-ordered events in a 
sequence $\tau$ 
that can be reversed by \sdpor \cite{abdulla2014optimal} to possibly obtain
a previously unexplored trace. \newline
$\ithbr{\tau}{e'}{e}$ if $\sw{\tau}{e'}{e}$ $\v$ $\dob{\tau}{e'}{e}$
\newline

\noindent
Let (ithb-irreversible) $\emithbi$ represent the $\emithb$-ordered events in a 
sequence $\tau$ 
that cannot be reversed by \sdpor \cite{abdulla2014optimal} to maintain
coherence. \newline
$\ithbi{\tau}{e'}{e}$ if $\ithb{\tau}{e'}{e}$ $\^$ $\nithbr{\tau}{e'}{e}$

\theorem{$\hb{\tau}{}{}$ for any sequence $\tau$ is valid}{appendix valid hb}

A happens-before assignment is valid if $\forall \tau$, $\hb{\tau}{}{}$ 
satisfies the following properties \cite{abdulla2014optimal}.
	
	\begin{enumerate}
		\item $\hb{\tau}{}{}$ is a partial order on $\events_\tau$, which 
			is included in $\ob{\tau}{}{}$.
			
		\item Events of each processing element are totally ordered, \ie
			$\hb{\tau}{\langle \p{t}, a', o', m', i \rangle}
			{\langle \p{t}, a, o, m, i+1 \rangle}$, whenever
			$\langle \p{t}, a, o, m, i+1 \rangle \in \events_\tau$.
			
		\item If $\tau'$ is a prefix of $\tau$, then $\hb{\tau'}{}{}$ and 
			$\hb{\tau}{}{}$ are the same on $\events_{\tau'}$
		
		\item Any linearization $\tau'$ of $\hb{\tau}{}{} \union 
			\rf{\tau}{}{}$ is an execution 
			sequence \st $\hb{\tau'}{}{} \union \rf{\tau'}{}{}$ = 
			$\hb{\tau}{}{} \union \rf{\tau}{}{}$ and 
			$\tau' \equi \tau$.
			
		\item If $\tau' \equi \tau$ then $\s{\tau'} = \s{\tau}$.
			
		\item If $\tau.\tau_1$ is an execution sequence, then $\tau' \equi
			\tau \iff \tau'.\tau_1 \equi \tau.\tau_1$
			
		\item If $\hb{\tau.e_1.e_2}{e_1}{e_2}$ and $\nhb{\tau.e_1.e_3}{e_1}{e_3}$
			then $\hb{\tau.e_1.e_3.e_2}{e_1}{e_2}$.	
	\end{enumerate}

\proof{
	\begin{itemize}
		\item Property 1,3 follow directly from construction of $\hb{\tau}{}{}$.
		
		\item Property 2 follows from the definition of $\po{\tau}{}{}$.
		
		\item Assume $\exists \tau_1$ \st 
			$\tau_1 \nequi \tau$ then $\exists e',e$ \st $\hb{\tau_1}{e}{e'}$
			but $\nhb{\tau}{e}{e'}$ $\implies$ $\hb{\tau_1}{}{} \neq 
			\hb{\tau}{}{}$. Thus, Property 4 follows from construction of 
			$\hb{\tau}{}{}$.
			
		\item $\tau' \equi \tau$ 
		
			$\implies \hb{\tau'}{}{} \union \rf{\tau'}{}{}$ = $\hb{\tau}{}{} 
			\union \rf{\tau}{}{}$
			
			$\implies \events_\tau' = \events_\tau\ \^\ \forall e \in 
			\events_\tau, val(\tau',e) = val(\tau,e)$.
			
			$\implies \s{\tau'} = \s{\tau}$ 
			
		\item $\tau' \equi \tau$ 
		
			$\implies \hb{\tau'}{}{} \union \rf{\tau'}{}{}$ = $\hb{\tau}{}{} 
			\union \rf{\tau}{}{}$
		
			$\implies \hb{\tau'}{}{} \union \rf{\tau'}{}{} \union \hb{\tau_1}{}{} 
			\union \rf{\tau_1}{}{}$ = $\hb{\tau}{}{} \union \rf{\tau}{}{} \union
			\hb{\tau_1}{}{} \union \rf{\tau_1}{}{}$
		
			$\implies \tau'.\tau_1 \equi \tau.\tau_1$ \newline
		
			And similarly, $\tau'.\tau_1 \equi \tau.\tau_1$
		
			$\implies \hb{\tau'}{}{} \union \rf{\tau'}{}{} \union \hb{\tau_1}{}{} 
			\union \rf{\tau_1}{}{}$ = $\hb{\tau}{}{} \union \rf{\tau}{}{} \union
			\hb{\tau_1}{}{} \union \rf{\tau_1}{}{}$
		
			$\implies \hb{\tau'}{}{} \union \rf{\tau'}{}{}$ = $\hb{\tau}{}{} 
			\union \rf{\tau}{}{}$
		
			$\implies \tau' \equi \tau$
		
		\item $\hb{\tau.e_1.e_2}{e_1}{e_2}$
		
			$\implies \po{\tau.e_1.e_2}{e_1}{e_2}$ or $\ithbr{\tau.e_1.e_2}{e_1}{e_2}$
			
			if $\po{\tau.e_1.e_2}{e_1}{e_2}$ then $\po{\tau.e_1.e_3.e_2}{e_1}{e_2}$
			(by definition of $\empo$)
			
			if $\ithbr{\tau.e_1.e_2}{e_1}{e_2}$ then $\ithbr{\tau.e_1.e_3.e_2}{e_1}{e_2}$
			(as $\nhb{\tau.e_1.e_3}{e_1}{e_3}$)
			
			Note that $\nithbi{\tau.e_1.e_2}{e_1}{e_2}$ because they are adjacent.
	\end{itemize}
}

\theorem{All \ourtechnique traces are coherent \cmodel traces.}{appendix coherence}

\noindent
The happens-before relation creates a partial order on the program events that
represents a trace. Thus alternatively, soundness can be stated as:
\textit{\ourtechnique's happens-before is a subset of
	\cmodel's happens-before} \ie $\forall \tau,\ \hb{\tau}{}{}\
\subseteq\ \chb{\tau}{}{}$ \newline

\proof{
\begin{enumerate}
	\item 
		Recall, (\hl{mo1}): $\forall w_1,w_2 \in \writes_\tau$, 
		$\chb{\tau}{w_1}{w_2} \implies \mo{\tau}{w_1}{w_2}$.
	
		Now, if $\hb{\tau}{w_1}{w_2}$
	
		$\implies \po{\tau}{w_1}{w_2} \v \ithbi{\tau}{w_1}{w_2}$
		
		$\po{\tau}{w_1}{w_2} \implies thr(shw(w_1)) = thr(shw(w_2)) = \tid{i}$ (say),
		$\^$ $\ob{\tid{i}}{shw(w_1)}{shw(w_2)}$
	
		$\ithbi{\tau}{w_1}{w_2} \implies \mhb{\tau}{w_1}{w_2}$	
		$\implies \ob{\tau}{shw(w_1)}{shw(w_2)}$ (by \hlref{shmo1})
		
		$\implies \mo{\tau}{w_1}{w_2}$ (using (\hlref{shmo}))
	
	\item 
		Recall, (\hl{mo2}): $\forall r_1,r_2 \in \reads_\tau,\ w_1 
		\in \writes_\tau, \chb{\tau}{r_1}{r_2}\ \^\
		\crf{\tau}{w_1}{r_1} \implies \crf{\tau}{w_1}{r_2}\ \v\
		(\exists w_2, \crf{\tau}{w_2}{r_2}\ \^\ 
		\mo{\tau}{w_1}{w_2})$.
		
		Now, $\hb{\tau}{r_1}{r_2}\ \^\ w_1 \neq w_2 \implies 
		\mo{\tau}{w_1}{w_2}$ (using rule (\hlref{shmo2}), (\hlref{shmo}))
		
	\item Recall, (\hl{mo3}): $\forall r_1 \in \reads_\tau,\ w_1 \in 
		\writes_\tau,\ \chb{\tau}{r_1}{w_1} \implies \exists 
		\mo{\tau}{w_2}{w_1} \st \crf{\tau}{w_2}{r_1}$.
		
		Now, if $\hb{\tau}{r_1}{w_1}$, $\exists w_2, \rf{\tau}{w_2}{r_1}$
		(by definition of (\hlref{shco})) 
		
		then by definition of $\rf{\tau}{}{}$, 
		
		\tab either $\ob{\tau}{shw(w_2)}{r_1} \implies 
		\ob{\tau}{\ob{\tau}{shw(w_2)}{r_1}}{shw(w_1)}$
		
		\tab or $thr(w_2) = thr(r_1) \implies \mhb{\tau}{\mhb{\tau}{w_2}{r_1}}{w_1}$
		(as $\hb{\tau}{r_1}{w_1} \implies \po{\tau}{r_1}{w_1} \v
		\ithbi{\tau}{r_1}{w_1}$)
		$\implies \ob{\tau}{shw(w_2)}{shw(w_1)}$ (by \hlref{shmo1})
		
		$\implies \mo{\tau}{w_2}{w_1}$ (using (\hlref{shmo}))
		
	\item Recall, (\hl{mo4}): $\forall w_1 \in \writes_\tau,\ r_1 \in \reads_\tau,
		\chb{\tau}{w_1}{r_1} \implies \crf{\tau}{w_1}{r_1} \v
		(\exists w_2, \crf{\tau}{w_2}{r_1}\ \^\ \mo{\tau}{w_1}{w_2})$.
	
		Now, $\hb{\tau}{w_1}{r_1} \implies \sw{\tau}{w_1}{r_1}\ \v\ 
		\dob{\tau}{w_1}{r_1}\ \v\ \po{\tau}{w_1}{r_1}\ \v\ \ithbi{\tau}{w_1}{r_1}$
		\hfill (by definition of $\hb{\tau}{}{}$)
		
		\tab $\sw{\tau}{w_1}{r_1} \implies w_1 = w_2$
		
		\tab $\dob{\tau}{w_1}{r_1} \implies \subseq{\langle w_2 \rangle}{\rs{\tau}{w_1}}
			\implies \mo{\tau}{w_1}{w_2}$ (by definition of $\rs{\tau}{w_1}$)
		
		\tab $\po{\tau}{w_1}{r_1}\ \v\ \ithbi{\tau}{w_1}{r_1} \implies 
			\mhb{\tau}{w_1}{r_1}$ then $\rf{\tau}{w_2}{r_1} \implies 
			\ob{\tau}{shw(w_1)}{shw(w_2)} \implies \mo{\tau}{w_1}{w_2}$
			(using (\hlref{shmo}), (\hlref{shmo1}))
			
		$\implies (w_1 = w_2$ \ie $\rf{\tau}{w_1}{r_1})\ \v \ \mo{\tau}{w_1}{w_2}$
		
	\item Recall, (\hl{to}): $\nexists sc_1, sc_2 \in \events_\tau^{(\sc)}$ $\st$ 
		$\cto{\tau}{sc_1}{sc_2}$ $\^$ $(\chb{\tau}{sc_2}{sc_1}$ $\v$ 
		$\cmo{\tau}{sc_2}{sc_1})$.
		
		Now, by definition $\to{\tau}{sc_1}{sc_2} \implies \ob{\tau}{sc_1}{sc_2}\ \v\ 
		\ob{\tau}{shw(sc_1)}{sc_2}\ \v \ \ob{\tau}{sc_1}{shw(sc_2)}\ \v\ 
		\ob{\tau}{shw(sc_1)}{shw(sc_2)}$
		
		$\implies \nhb{\tau}{sc_2}{sc_1}$
	
	\item Recall, (\hl{co}): $val(\tau,r_1) = v$ for $r_1 \in \reads_\tau$ then 
		$\exists w_1 \in \writes_\tau \st val(\tau,w_1) = v \^ 
		\nchb{\tau}{r_1}{w_1}$
		
		Now, $val(\tau,w_1) = v = val(\tau,r_1) \implies \ob{\tau}{w_1}{r_1}$
		
		$\implies \nhb{\tau}{r_1}{w_1}$  (using (\hlref{shco}))
\end{enumerate}

Proof statements 1. to 6. $\implies$ all valid \ourtechnique traces are valid \cmodel
traces. Implying, that our technique does not explore a non-\cmodel behavior.
}

\newpage
\section{Proof of Soundness} \label{proof: soundness}
\theorem{\Pretransformation (\program to \programhat) is semantic preserving.}
{appendix Phat semantic preserving}

\proof{
	Consider the sequence of events $\tid{i}{:}\tau$ \ie the sequence of
	events to be executed by thread $\tid{i}$ of original program \program.
	Further, consider the reordered sequence of events $\tid{i}{:}\tau$
	to be executed by the corresponding thread $\tid{i}$ of transformed
	program \programhat.
	
	As discussed in \refsection{sec:mca}, to ensure a semantic 
	preserving reordering from \program to \programhat the following 
	must hold for each thread $\tid{i}$ of \program, 
	
\begin{itemize}[\IEEEsetlabelwidth{spr1}]
	\item  [spr1] $\events_{\tid{i}{:}\tau'}
	= \events_{\tid{i}{:}\tau}$ (the event sets are the same)
	\item [spr2] each
	read event of $\tid{i}{:}\tau$ must have the same set of writes to
	read from in $\tid{i}{:}\tau$ as well as $\tid{i}{:}\tau'$ (we call it {\em
		thread  semantics})
	\item [spr3] $\tid{i}{:}\tau'$ preserves  
	the order of updates and accesses of each shared variable
	with respect to $\tid{i}{:}\tau$ ({\em coherence-per-location}).
\end{itemize}
	
	\begin{enumerate}
		\item [spr1] By definition of \pretransformation discussed in 
			\refsection{sec:algo}, $\events_{\tid{i}{:}\tau'}
			= \events_{\tid{i}{:}\tau}$.
			
		\item [spr2] $\forall$ events $e',e$ to be executed by thread $\tid{i}$ 
			\st $obj(e') = obj(e)\ \v\ \dep{e'}{e}$,
			$\ob{\tid{i}{:}\tau}{e'}{e}\ \^\ \ob{\tid{i}{:}\tau'}{e'}{e}$
			(by definition of \pretransformation).
			
			Thus, $\forall e_r \in \reads$ of $\p{t_i}$, 
			$\pw{\tid{i}{:}\tau}{e_r}$ = $\pw{\tid{i}{:}\tau'}{e_r}$.
			
			(Where, $\pw{\tid{i}{:}\tau}{e}$ represents the previous write event to the
			event $e$ in thread $t$ \ie $e_w \in \writes = \pw{\tid{i}{:}\tau}{e}$ if 
			$thr(e_w) = thr(e) = t, \ob{\tid{i}{:}\tau}{e_w}{e}$ and $\nexists e_w' 
			\in \writes \st thr(e_w') = t \^ \ob{\tid{i}{:}\tau}{e_w < e_w'}{e}$.)
			
			$\implies$ $\tid{i}{:}\tau'$ preserves the sequential semantics of 
			$\tid{i}{:}\tau$.
			
		\item [spr3] $\forall o \in \objects,\ \forall e',e \in \events$ of $\tid{i}{:}\tau$
			\st $obj(e') = obj(e) = o$, if $\ob{\tid{i}{:}\tau}{e'}{e}$ then
			$\ob{\tid{i}{:}\tau'}{e'}{e}$
			(by definition of \pretransformation).
			
			$\implies$ $\tid{i}{:}\tau'$ preserves {\em coherence-per-location}.
	\end{enumerate}
	Thus, \pretransformation (\program to \programhat) is semantic preserving. 
}

\lemma{If $e',e$ can reorder under \mca then their effects can 
	interleave under \ourtechnique}{mca reordering supported}

\proof{
	For all events $e',e$ such that $\eventreorder{e'}{e}$, there are four valid 
	possibilities considered below.
	
	\begin{enumerate}
		\item $e',e \in \reads$: 
		
		$\forall \tau \st e',e \in \reads_\tau, \exists w',w \in \writes_\tau
		\st \rf{\tau}{w'}{e'},\ \rf{\tau}{w}{e}$.
		
		$(\eventreorder{e'}{e}) \implies obj(e') \neq obj(e)$
		
		$\implies (obj(w') = obj(e')) \neq (obj(w) = obj(e))$
		
		Further, $obj(w') \neq obj(w) \implies thr(shw(w')) \neq thr(shw(w))$
		
		$\implies$ $shw(w')$ and $shw(w)$ can be interleaved.
		
		\item $e' \in \reads, e \in \writes$
		
		$(\eventreorder{e'}{e}) \implies obj(e') \neq obj(e)$ and 
		$\neg \dep{e'}{e}$
		
		$\implies \ob{\tid{i}{:}\tau}{e'}{e}$ but $\ob{\tid{i}{:}\tau'}{e}{e'}$
		and $shw(e)$ and $e'$ can be interleaved.
		
		\item $e' \in \writes, e \in \reads$
		
		$(\eventreorder{e'}{e}) \implies obj(e') \neq obj(e)$ and 
		$thr(shw(e')) \neq thr(e)$
		
		$\implies$ $shw(e')$ and $e$ can be interleaved.
		
		\item $e',e \in \writes$
		
		$(\eventreorder{e'}{e}) \implies obj(e') \neq obj(e) 
		\implies thr(shw(e')) \neq thr(shw(e))$
		
		$\implies$ $shw(e')$ and $shw(e)$ can be interleaved.
	\end{enumerate}
}

\lemma{\textit{\ourtechnique $\subseteq$ \cmodel under \mca}: 
	All \ourtechnique traces are \cmodel traces valid over \mca.}
{appendix our subset mca}

\proof{
$\forall e',e \st \neg(\eventreorder{e'}{e}) \implies$
$e',e$ have distinct local and shared variables

$\implies dep(e',e) \implies$ if $\ob{\tid{i}{:}\tau}{e'}{e}$ then
$\ob{\tid{i}{:}\tau'}{e'}{e}$ 

Further, if $e',e \in \writes$ then $thr(shw(e')) = thr(shw(e)) = t$
then $\ob{\tid{i}{:}\tau}{shw(e')}{shw(e)}$; (by construction of shadow-threads) and,

if $e'\in \writes, e \in \reads$ then $e$ can observe effect of 
$e'$ even when $\ob{\tau}{e}{shw(e')}$ (by definition of 
$\emrf$)

and, if $\noup{e',e}$ then if $\ob{\tid{i}{:}\tau}{e'}{e}$ then
$\ob{\tid{i}{:}\tau'}{e'}{e}$ 

hence, reordering restricted by \cmodel over \mca is also 
restricted by \ourtechnique.\hfill ...(i) \newline

Since, \ourtechnique traces are coherent \cmodel traces
(Theorem \ref{thm:appendix coherence}), the lemma can be reduced to:
if a trace is allowed by \ourtechnique it must be allowed under
\mca. \newline

\pretransformation is semantic preserving (Theorem 
\ref{thm:appendix Phat semantic preserving}) 

Apart from \pretransformation, \ourtechnique can reorder writes
down with shadow-writes.
However, $\forall e_w \in \writes, e_r \in \reads \st 
\ob{\tid{i}{:}\tau}{e_w}{e_r}$, $e_r$ can observe $e_w$ 
(using rule (\hlref{shco}));and,

$\forall e_{w1}, e_{w2} \in \writes \st 
\ob{\tid{i}{:}\tau'}{e_{w1}}{e_{w2}}$, 
$\ob{\tau}{shw(e_{w1})}{shw(e_{w2})}$ for all 
sequences $\tau$. 

hence, reordering allowed by \ourtechnique is allowed under \mca 
\hfill ...(ii)\newline

$\forall e_w \in \writes$ along with $shw(e_w)$ perform 
\hlref{(w-issue)} and \hlref{(w-update)}

while $shw(e_w)$ delays update of $e_w$ to memory but still
$e_r \in \reads \st \ob{\tid{i}{:}\tau'}{e_w}{e_r}$ can observe $e_w$
(using rule (\hlref{shco})) and $\forall e_{w2} \in \writes \st
\ob{\tid{i}{:}\tau'}{e_w}{e_{w2}}$, $\ob{\tau}{shw(e_w)}{shw(e_{w2})}$
for all sequences $\tau$ (by construction of shadow-threads)

Thus, $e_w \in \writes$ along with $shw(e_w)$ perform 
\hlref{(w-issue)} and \hlref{(w-update)} while preserving 
sequential semantics and coherence per location.

$\forall e_r \in \reads$ perform \hlref{(r-shared)}.

Lastly, \sdpor algorithm \cite{abdulla2014optimal} ensures \hlref{(parcom)}.

Thus, sequences of \ourtechnique have equivalent \mca 
execution sequences. \hfill ...(iii)\newline

Hence proved from (i), (ii) and (iii).
}

\lemma{\cmodel under \mca $\subseteq$ \ourtechnique: 
	All \cmodel traces valid over \mca are valid \ourtechnique traces}
{appendix mca subset our}

\proof{
	If $\ob{\tid{i}{:}\tau}{e'}{e}$ and $\ob{\tid{i}{:}\tau'}{e'}{e}$ then 
	$\dep{e'}{e} \v \noup{e',e}$
	
	$\dep{e'}{e} \implies \neg(\eventreorder{e'}{e})$

	$\noup{e',e} \implies$ reordering of $e'$ and $e$ is not supported
	by \cmodel.
	
	Further, if $\dep{e'}{e}$ then if $e',e \in \writes$ then, 
	$\ob{\tau}{shw(e')}{shw(e)}$ ($e \in \writes$), 
	and if $e' \in \writes, e \in \reads$ then $e_r$ can observe $e_w$
	(using (\hlref{shco})).
	
	Hence, reordering restricted by \ourtechnique is also restricted 
	by \cmodel under \mca \hfill ...(i) \newline
	
	If $e',e$ can reorder under \mca then their effects can 
	interleave under \ourtechnique Lemma 
	\ref{lem:mca reordering supported}	
	
	Hence, reordering allowed under \mca are allowed by 
	\ourtechnique \hfill ...(ii)\newline
	
	\ourtechnique traces are coherent \cmodel traces (Theorem 
	\ref{thm:appendix coherence}).
	
	Thus, the lemma can be reduced to: if a behavior is allowed under
	\mca then it is also allowed by \ourtechnique.
	
	\hlref{(r-shared)} is performed by $e \in \reads$.
	
	\hlref{(w-issue)} and \hlref{(w-update)} is performed by
	$e \in \writes$ and corresponding $shw(e)$.
	
	\hlref{(parcom)} is ensured by \sdpor algorithm.
	
	Hence, sequences produced by \mca semantics can be 
	replicated by \ourtechnique traces. \hfill...(iii)
	
	Hence proved using (i), (ii) and (iii).
}

\theorem{\ourtechnique traces are equivalent to \cmodel traces valid over \mca}
{appendix traces eq}

\proof{
	\ourtechnique traces $\subseteq$ \mca traces. (using Lemma 
	\ref{lem:appendix our subset mca}).
	
	\cmodel traces valid over \mca $\subseteq$ \ourtechnique traces. 
	(using Lemma \ref{lem:appendix mca subset our}).
	
	Thus, \ourtechnique traces are equivalent to \cmodel traces over \mca.
}

\end{appendices}
\end{document}